\documentclass[useAMS, usenatbib]{mn2e}

\usepackage[breaklinks,colorlinks,citecolor=blue]{hyperref}
\usepackage{graphicx, bm, amssymb, xcolor}
\usepackage{verbatim}
\usepackage[all]{hypcap}
\newcommand{\noopsort}[1]{}

\newcommand{\slug}{\texttt{slug}}
\newcommand{\Slug}{\texttt{Slug}}
\newcommand{\bp}{\texttt{bayesphot}}
\newcommand{\Bp}{\texttt{Bayesphot}}
\newcommand{\cs}{\texttt{cluster\_slug}}

\newcommand{\sfrslug}{\texttt{SFR\_slug}}
\newcommand{\Sfrslug}{\texttt{SFR\_slug}}
\newcommand{\clusterslug}{\texttt{cluster\_slug}}

\newcommand{\cloudyslug}{\texttt{cloudy\_slug}}
\newcommand{\Cloudyslug}{\texttt{Cloudy\_slug}}
\newcommand{\slugpy}{\texttt{slugpy}}

\newcommand{\vech}{\mathbf{h}}
\newcommand{\vecx}{\mathbf{x}}
\newcommand{\vecy}{\mathbf{y}}
\newcommand{\vecs}{\boldsymbol{\sigma}_{\mathbf{y}}}
\newcommand{\vecyo}{\mathbf{y}_{\mathrm{obs}}}

\newcommand{\apj}{ApJ}
\newcommand{\apjl}{ApJ}
\newcommand{\apjs}{ApJS}
\newcommand{\aj}{AJ}
\newcommand{\aaps}{A\&AS}
\newcommand{\araa}{ARA\&A}
\newcommand{\mnras}{MNRAS}
\newcommand{\aap}{A\&A}

\newcommand{\jacm}{J.~Assoc.~Comput.~Mach.}

\newcommand{\pasp}{Proc.~Astron.~Soc.~Pac.}
\newcommand{\rmxaa}{Rev.~Mex.~Astron.~Astrop.}
\newcommand{\nar}{New~Astron.~Rev.}

\newcommand{\red}[1]{#1}

\begin{document}

\title[SLUG III]{SLUG -- Stochastically Lighting Up Galaxies. III: A Suite of Tools for Simulated Photometry, Spectroscopy,  and Bayesian Inference with Stochastic Stellar Populations}

\author[Krumholz et al.]{Mark R. Krumholz$^1$\thanks{mkrumhol@ucsc.edu},
Michele Fumagalli$^{2,3}$\thanks{michele.fumagalli@durham.ac.uk},
Robert L.~da Silva$^1$\thanks{rdasilva.astro@gmail.com},
\newauthor
Theodore Rendahl$^1$,
Jonathan Parra$^1$
\\ \\
$^1$Department of Astronomy \& Astrophysics, University of California, Santa 
Cruz, CA 95064 USA\\
$^2$ Institute for Computational Cosmology and Centre for Extragalactic Astronomy, 
     Department of Physics, \\  ~~Durham University, South Road, Durham, DH1 3LE, UK \\
$^3$Carnegie Observatories, 813 Santa Barbara Street, Pasadena, CA 91101, USA \\
}

\date{\today}

\pagerange{\pageref{firstpage}--\pageref{lastpage}} \pubyear{2015}

\maketitle

\label{firstpage}

\begin{abstract} 
Stellar population synthesis techniques for predicting the observable light emitted by a stellar population have extensive applications in numerous areas of astronomy. However, accurate predictions for small populations of young stars, such as those found in individual star clusters, star-forming dwarf galaxies, and small segments of spiral galaxies, require that the population be treated stochastically. Conversely, accurate deductions of the properties of such objects also requires consideration of stochasticity. Here we describe a comprehensive suite of modular, open-source software tools for tackling these related problems. These include: a greatly-enhanced version of the \slug\ code introduced by \citet{da-silva12a}, which computes spectra and photometry for stochastically- or deterministically-sampled stellar populations with nearly-arbitrary star formation histories, clustering properties, and initial mass functions; \cloudyslug, a tool that automatically couples \slug-computed spectra with the \texttt{cloudy} radiative transfer code in order to predict stochastic nebular emission; \bp, a general-purpose tool for performing Bayesian inference on the physical properties of stellar systems based on unresolved photometry; and \cs\ and \sfrslug, a pair of tools that use \bp\ on a library of \slug\ models to compute the mass, age, and extinction of mono-age star clusters, and the star formation rate of galaxies, respectively. The latter two tools make use of an extensive library of pre-computed stellar population models, which are included the software.  The complete package is available at \url{http://www.slugsps.com}.
\end{abstract}

\section{Introduction}
\label{sec:intro}

Stellar population synthesis (SPS) is a critical tool that allows us to link the observed light we receive from unresolved stellar populations to the physical properties (e.g.~mass, age) of the emitting stars. Reflecting this importance, over the years numerous research groups have written and distributed SPS codes such as \texttt{starburst99} \citep{leitherer99a, vazquez05a}, \texttt{fsps} \citep{conroy09a, conroy10b}, \texttt{pegase} \citep{fioc97a}, and \texttt{galaxev} \citep{bruzual03a}. All these codes perform essentially the same computation. One adopts a star formation history (SFH) and an initial mass function (IMF) to determine the present-day distribution of stellar masses and ages. Next, using a set of stellar evolutionary tracks and atmospheres that give the luminosity (either frequency-dependent or integrated over some filter) for a star of a particular mass and age, one integrates the stellar luminosity weighted by the age and mass distributions. These codes differ in the range of functional forms they allow for the IMF and SFH, and the evolutionary tracks and model atmospheres they use, but the underlying computation is much the same in all of them. While this approach is adequate for many applications, it fails for systems with low masses and star formation rates (SFRs) because it implicitly assumes that the stellar mass and age distributions are well-sampled. This is a very poor assumption both in star-forming dwarf galaxies and in resolved segments of larger galaxies.

Significantly less work has been done in extending SPS methods to the regime where the IMF and SFH are not well-sampled. There are a number of codes available for simulating a simple stellar population (i.e., one where all the stars are the same age, so the SFH is described by a $\delta$ distribution) where the IMF is not well sampled \citep{maiz-apellaniz09a, popescu09a, popescu10a, popescu10b, fouesneau10a, fouesneau12a, fouesneau14a, anders13a, de-meulenaer13a, de-meulenaer14a, de-meulenaer15a}, and a great deal of analytic work has also been performed on this topic \citep{cervino03a, cervino04a, cervino06a} -- see \citet{cervino13a} for a recent review. However, these codes only address the problem of stochastic sampling of the IMF; for non-simple stellar populations, stochastic sampling of the SFH proves to be a more important effect \citep{fumagalli11a, da-silva14b}.

To handle this problem, we introduced the stochastic SPS code \slug~\citep{da-silva12a}, which includes full stochasticity in both the IMF and the SFH. Crucially, \slug\ properly handles the clustered nature of star formation \citep[e.g.,][]{lada03a, krumholz14c}. \red{This has two effects. First, clustering itself can interact with IMF sampling so that the total mass distribution produced by a clustered population is not identical to that of a non-clustered population drawn from the same underlying IMF. Second and perhaps more importantly, clustering} causes large short-timescale fluctuations in the SFR even in galaxies whose mean SFR averaged over longer timescales is constant. Since its introduction, this code has been used in a number of applications, including explaining the observed deficit of H$\alpha$ emission relative to FUV emission in dwarf galaxies \citep{fumagalli11a, andrews13a, andrews14a}, quantifying the stochastic uncertainties in SFR indicators \citep{da-silva14b}, analyzing the properties of star clusters in dwarf galaxies in the ANGST survey \citep{cook12a}, and analyzing Lyman continuum radiation from high-redshift dwarf galaxies \citep{forero-romero13a}. The need for a code with stochastic capabilities is likely to increase in the future, as studies of local galaxies such as PHAT \citep{dalcanton12a}, HERACLES \citep{leroy13a}, and LEGUS \citep{calzetti15a}, and even studies in the high redshift universe \citep[e.g.,][]{jones13a, jones13b} increasingly push to smaller spatial scales and lower rates of star formation, where stochastic effects become increasingly important.

In this paper we describe a major upgrade and expansion of \slug, intended to make it a general-purpose solution for the analysis of stochastic stellar populations. This new version of \slug~allows essentially arbitrary functional forms for both the IMF and the SFH, allows a wide range of stellar evolutionary tracks and atmosphere models, and can output both spectroscopy and photometry in a list of $>100$ filters. It can also include the effects of reprocessing of the light by interstellar gas and by stochastically-varying amounts of dust, and can interface with the \texttt{cloudy} photoionization code \citep{ferland13a} to produce predictions for stochastically-varying nebular emission. Finally, we have coupled \slug~to a new set of tools for solving the ``inverse problem" in stochastic stellar population synthesis: given a set of observed photometric properties, infer the posterior probability distribution for the properties of the underlying stellar population, in the case where the mapping between such properties and the photometry is stochastic and therefore non-deterministic \citep[e.g.][]{da-silva14b}. The full software suite is released under the GNU Public License, and is freely available from \url{http://www.slugsps.com}.\footnote{As of this writing \url{http://www.slugsps.com} is hosted on Google Sites, and thus is inaccessible from mainland China. Chinese users can access the source code from \url{https://bitbucket.org/krumholz/slug2}, and should contact the authors by email for the ancillary data products.}

In the remainder of this paper, we describe \slug~and its companion software tools in detail (\autoref{sec:slug}), and then provide a series of demonstrations of the capabilities of the upgraded version of the code (\autoref{sec:applications}). We end with a summary and discussion of future prospects for this work (\autoref{sec:summary}).

\section{The \texttt{SLUG} Software Suite}
\label{sec:slug}

\red{The \slug~software suite is a collection of tools designed to solve two related problems. The first is to determine the probability distribution function (PDF) of observable quantities (spectra, photometry, etc.) that are produced by a stellar population characterized by a specified set of physical parameters (IMF, SFH, cluster mass function, and an array of additional ancillary inputs). This problem is addressed by the core \slug~code (\autoref{ssec:slugcore}) and its adjunct \cloudyslug~(\autoref{sec:cldy}), which perform Monte Carlo simulations to calculate the distribution of observables. The second problem is to use those PDFs to solve the inverse problem: given a set of observed properties, what should we infer about the physical properties of the stellar population producing those observables? The \bp~package provides a general solution to this problem (\autoref{ssec:bp}), and the \sfrslug~and \cs~packages specialize this general solution to the problem of inferring star formation rates for continuously star-forming galaxies, and masses, ages, and extinctions from simple stellar populations, respectively (\autoref{ssec:sfr_cluster_slug}). The entire software suite is available from \url{http://www.slugsps.com}, and extensive documentation is available at \url{http://slug2.readthedocs.org/en/latest/}.}

\subsection{\slug: A Highly Flexible Tool for Stochastic Stellar Population Synthesis}
\label{ssec:slugcore}

The core of the software suite is the stochastic stellar population synthesis code \slug. The original \slug\ code is described in \citet{da-silva12a}, \red{but the code described here is a complete re-implementation with greatly expanded capabilities. The code operates in two main steps; first it generates the spectra from the stellar population (\autoref{sssec:generate}) and then it post-processes the spectra to include the effects of nebular emission and dust absorption, and to produce photometric predictions as well as spectra (\autoref{sssec:ppspectra}). In this section we limit ourselves to a top-level description of the physical model, and provide some more details on the numerical implementation in Appendices \ref{app:implementation}, \ref{app:nebuladust}, and \ref{app:software}.}

\subsubsection{\red{Generating the Stellar Population}}
\label{sssec:generate}

\red{\Slug\ can simulate both simple stellar populations (SSPs, i.e., ones where all the stars are coeval) and composite ones. We begin by describing the SSP model, since composite stellar populations in \slug\ are built from collections of SSPs. For an SSP, the stellar population is described by an age, a total mass, and an IMF. \Slug\ allows nearly-arbitrary functional forms for the IMF, and is not restricted to a set of predetermined choices; see Appendix \ref{app:pdfs} for details.}

\red{In non-stochastic SPS codes, once an age, mass, and IMF are chosen, the standard procedure is to use a set of stellar tracks and atmospheres to predict the luminosity (either frequency-dependent, or integrated over a specified filter) as a function of stellar mass and age, and to integrate the mass-dependent luminosity multiplied by the stellar mass distribution to compute the output spectrum at a specified age. \Slug\ adds an extra step: instead of evaluating an integral, it directly draws stars from the IMF until the desired total mass has been reached.} As emphasized by a number of previous authors \red{\citep[e.g.][]{weidner06a, haas10a, cervino13a, popescu14a}}, when drawing a target mass $M_{\rm target}$ rather than a specified number of objects from a PDF, one must also choose a sampling method to handle the fact that, in general, it will not be possible to hit the target mass exactly. \red{Many sampling procedures have been explored in the literature, and \slug\ provides a large number of options, as described in Appendix \ref{app:pdfs}.}

\red{Once a set of stars is chosen, \slug\ proceeds much like a conventional SPS code. It uses a chosen set of tracks and atmospheres to determine the luminosity of each star, either wavelength-dependent or integrated over one or more photometric filters, at the specified age. It then sums over all stars to determine the integrated light output. Details of the available sets of tracks and atmospheres, and \slug's method for interpolating them, are provided in Appendix \ref{app:tracks}.}

\red{Composite stellar populations in \slug\ consist of a collection of ``star clusters", each consisting of an SSP, plus a collection of ``field stars" whose ages are distributed continuously. In practice, this population is described by four parameters beyond those that describe SSPs: the fraction of stars formed in clusters as opposed to the field $f_c$, the cluster mass function (CMF), cluster lifetime function (CLF), and star formation history (SFH). As with the IMF, the latter three are distributions, which can be specified using nearly arbitrary functional forms and a wide range of sampling methods as described in Appendix \ref{app:pdfs}. For a simulation with a composite stellar population, \slug\ performs a series of steps: (1)} at each user-requested output time\red{\footnote{\red{\Slug~can either output results at specified times, or the user can specify a distribution from which the output time is to be drawn. The latter capability is useful when we wish to sample a distribution of times continuously so that the resulting data set can be used to infer ages from observations -- see \autoref{ssec:sfr_cluster_slug}.}}}, \slug\ uses the SFH and the cluster fraction $f_c$ to compute the additional stellar mass expected to have formed in clusters and out of clusters (in ``the field'') since the previous output; \red{(2)} for the non-clustered portion of the star formation, \slug\ draws the appropriate mass in individual stars from the IMF, while for the clustered portion it draws a cluster mass from the CMF, and then it fills each cluster with stars drawn from the IMF; \red{(3)} each star and star cluster formed is assigned an age between the current output and the previous one, selected to produce a realisation of the input SFH\red{\footnote{\red{One subtlety to note here is that the choice of output grid can produce non-trivial modifications of the statistical properties of the output in cases where the expected mass of stars formed is much smaller than the maximum possible cluster mass. For example, consider a simulation with a constant SFR and an output grid chosen such that the expected mass of stars formed per output time is $10^4$ $M_\odot$. If the sampling method chosen is \texttt{STOP\_NEAREST} (see Appendix \ref{app:pdfs}), the number of $10^6$ $M_\odot$ clusters formed will typically be smaller than if the same SFH were used but the outputs were spaced 100 times further apart, giving an expected mass of $10^6$ $M_\odot$ between outputs.}}}; \red{(4)} each cluster that is formed is assigned a lifetime drawn from the CLF, which is the time at which the cluster is considered dispersed.

The end result of this procedure is that, at each output time, \slug\ has constructed a ``galaxy'' consisting of a set of star clusters and field stars, each with a known age. Once the stellar population has been assembled, the procedure for determining spectra and photometry is simply a sum over the various individual simple populations. In addition to computing the integrated spectrum of the entire population, \slug\ can also report individual spectra for each cluster that has not disrupted (i.e., where the cluster age is less than the value drawn from the CLF for that cluster). Thus the primary output consists of an integrated \red{monochromatic} luminosity $L_\lambda$ for the entire stellar population, and a value $L_{\lambda,i}$ for the $i$th remaining cluster, at each output time.

\red{Finally, we note that \slug\ is also capable of skipping the sampling procedure and evaluating the light output of stellar populations by integrating over the IMF and SFH, exactly as in a conventional, non-stochastic SPS code. In this case, \slug\ essentially emulates \texttt{starburst99} \citep{leitherer99a, vazquez05a}, except for subtle differences in the interpolation and numerical integration schemes used (see Appendix \ref{app:tracks}). \Slug~can also behave ``semi-stochastically", evaluating stars' contribution to the light output using integrals to handle the IMF up to some mass, and using stochastic sampling at higher masses.}

\subsubsection{Post-Processing the Spectra}\label{sssec:ppspectra}

Once \slug\ has computed a spectrum for a \red{simple or composite stellar population}, it can perform three additional post-processing steps. First, it can provide an approximate spectrum after the starlight has been processed by the surrounding H~\textsc{ii} region. In this case, the nebular spectrum is computed for an isothermal, constant-density H~\textsc{ii} region, within which it is assumed that \red{He} is all singly-ionized. Under these circumstances, the photoionized volume $V$, electron density $n_e$, and hydrogen density $n_{\rm H}$ obey the relation 
\begin{equation}
\phi Q(\mathrm{H}^0) = \alpha^{\mathrm{(B)}}(T) n_e n_{\mathrm{H}} V
\end{equation}
where
\begin{equation}
Q(\mathrm{H}^0) = 
%\int_0^{hc/I(\mathrm{H}^0)} \frac{L_\lambda}{hc/\lambda} \, d\lambda
\red{\int_{h\nu = I(\mathrm{H}^0)}^\infty \frac{L_\nu}{h\nu} \, d\nu}
\end{equation}
is the hydrogen-ionizing luminosity, \red{$L_\nu = \lambda^2 L_\lambda/c$ is the luminosity per unit frequency,} $I(\mathrm{H}^0)=13.6$ eV is the ionization potential of neutral hydrogen, \red{$n_e = (1 + x_{\mathrm He}) n_{\mathrm{H}}$, $x_{\mathrm{He}}$ is the He abundance relative to hydrogen (assumed to be 0.1),} $\phi$ is the fraction of H-ionizing photons that are absorbed by hydrogen atoms \red{within the H~\textsc{ii} region rather than escaping or being absorbed by dust}, and $\alpha^{\mathrm{(B)}}(T)$ is the temperature-dependent case B recombination coefficient. \red{The nebular luminosity can then be written as
\begin{equation}
L_{\lambda,\mathrm{neb}} = \gamma_{\mathrm{neb}} n_e n_{\mathrm{H}} V = \gamma_{\mathrm{neb}} \phi \frac{Q(\mathrm{H}^0)}{\alpha^{\mathrm{(B)}}(T)},
\end{equation}
where $\gamma_{\mathrm{neb}}$ is the wavelength-dependent nebular emission coefficient. Our calculation of this quantity includes the following processes: free-free and bound-free emission arising from electrons interacting with H$^+$ and He$^+$, two-photon emission from neutral hydrogen in the $2s$ state, \red{H} recombination lines\red{, and non-H lines computed approximately from tabulated values}. Full details on the method by which we perform this calculation are given in Appendix \ref{app:nebuladust}. The composite spectrum after nebular processing is zero at wavelength shorter than 912 \AA~(corresponding to the ionization potential of hydrogen), and the sum of intrinsic stellar spectrum $L_{\lambda}$ and the nebular spectrum $L_{\lambda,\mathrm{neb}}$ at longer wavelengths. \Slug~reports this quantity both cluster-by-cluster and for the galaxy as a whole.}

Note that the treatment of nebular emission included in \slug\ is optimized for speed rather than accuracy, and will produce significantly less accurate results than \cloudyslug\ (see \autoref{sec:cldy}). \red{The major limitations of the built-in approach are: (1) because \slug's built-in calculation relies on pre-tabulated values for the metal line emission and assumes either a constant or a similarly pre-tabulated temperature (which significantly affects the continuum emission), it misses the variation in emission caused by the fact that H~\textsc{ii} regions exist at a range of densities and ionization parameters, which in turn induces substantial variations in their nebular emission \citep[e.g.][]{yeh13a, verdolini13a}; (2) \slug's built-in calculation assumes a uniform density H~\textsc{ii} region, an assumption that will fail at high ionizing luminosities and densities due to the effects of radiation pressure \citep{dopita02a, krumholz09a, draine11a, yeh12a, yeh13a}; (3) \slug's calculation correctly captures the effects of stochastic variation in the total ionizing luminosity, but it does not capture the effects of variation in the spectral shape of the ionizing continuum, which preliminary testing suggests can cause variations in line luminosities at the few tenths of a dex level even for fixed total ionizing flux.}
\red{The reason for accepting these limitations is that the assumptions that cause them also make it possible to express the nebular emission in terms of a few simple parameter that can be pre-tabulated, reducing the computational cost of evaluating the nebular emission by orders of magnitude compared to a fully accurate calculation with \cloudyslug.}

The second post-processing step available is that \slug\ can apply extinction in order to report an extincted spectrum, both for the pure stellar spectrum and for the nebula-processed spectrum. Extinction can be either fixed to a constant value for an entire simulation, or it can be drawn from a specified PDF. In the latter case, for simulations of composite stellar populations, every cluster will have a different extinction. \red{More details are given in Appendix \ref{app:nebuladust}.}

As a third and final post-processing step, \slug\ can convolve all the spectra it computes with one or more filter response functions in order to predict photometry. \Slug\ includes the large list of filter response functions maintained as part of \texttt{FSPS} \citep{conroy10a, conroy10b}, as well as a number of \textit{Hubble Space Telescope} filters\footnote{These filters were downloaded from \url{http://www.stsci.edu/~WFC3/UVIS/SystemThroughput/} and \url{http://www.stsci.edu/hst/acs/analysis/throughputs} for the UVIS and ACS instruments, respectively.} not included in \texttt{FSPS}; at present, more than 130 filters are available. As part of this calculation, \slug\ can also output the bolometric luminosity, and the \red{photon} luminosity in the H\red{$^0$}, He$^0$, and He$^+$ ionizing continua.

\subsection{\texttt{cloudy\_slug}: Stochastic Nebular Line Emission}
\label{sec:cldy}

In broad outlines, the \cloudyslug\ package is a software interface that automatically takes spectra generated by \slug\ and uses them as inputs to the \texttt{cloudy} radiative transfer code \citep{ferland13a}. \Cloudyslug\ then extracts the continuous spectra and lines returned by \texttt{cloudy}, convolves them with the same set of filters as in the original \slug\ calculation in order to predict photometry. \red{The user can optionally also examine all of \texttt{cloudy}'s detailed outputs. As with the core \slug~code, details on the software implementation are given in Appendix \ref{app:software}.}

When dealing with composite stellar populations, calculating the nebular emission produced by a stellar population requires making some physical assumptions about how the stars are arranged, and \cloudyslug\ allows two extreme choices that should bracket reality. One extreme is to assume that all stars \red{are concentrated in a single point of ionizing radiation at the center of a single H~\textsc{ii} region}. We refer to this as integrated mode, and in this mode the only free parameters to be specified by a user are the chemical composition of the gas into which the radiation propagates and its starting density.

The opposite assumption, which we refer to as cluster mode, is that every star cluster is surrounded by its own H~\textsc{ii} region, and that the properties of these regions are to be computed individually \red{using the stellar spectrum of each driving cluster} and only then summed to produce a composite output spectrum. \red{At present cluster mode does not consider contributions to the nebular emission from field stars, and so should not be used when the cluster fraction $f_c < 1$.} In \red{the cluster} case, the H~\textsc{ii} regions need not all have the same density or radius; indeed, one would expect a range of both properties to be present, since not all star clusters have the same age or ionizing luminosity. We handle this case using a simplified version of the H~\textsc{ii} population synthesis model introduced by \citet{verdolini13a} and \citet{yeh13a}. The free parameters to be specified in this model are the chemical composition and the density in the ambient neutral ISM around each cluster. For an initially-neutral region of uniform hydrogen number density $n_{\rm H}$, the radius of the H~\textsc{ii} at a time $t$ after it begins expanding is well-approximated by \citep{krumholz09d}
\begin{eqnarray}
r_{\rm II} & = & r_{\rm ch} \left(x_{\rm rad}^{7/2} + x_{\rm gas}^{7/2}\right)^{2/7} \\
x_{\mathrm{rad}} &= & \left(2\tau^2\right)^{1/4} \\
x_{\mathrm{gas}} &= & \left(\frac{49}{36}\tau^2\right)^{2/7} \\
\tau &= & \frac{t}{t_{\mathrm{ch}}} \\
r_{\mathrm{ch}} & = & \frac{\alpha^{\mathrm{(B)}}_4}{12\pi\phi_{\rm dust}} \left(\frac{\epsilon_0}{2.2 k_B T_{\mathrm{II}}}\right)^2
f_{\mathrm{trap}}^2 \frac{\psi^2 Q(\mathrm{H}^0)}{c^2} \\
t_{\mathrm{ch}} & = & \left(\frac{4\pi \mu m_{\mathrm{H}} n_{\mathrm{H}} c r_{\mathrm{ch}}^4}{3 f_{\mathrm{trap}}
   Q(\mathrm{H}^0) \psi I(\mathrm{H}^0)}\right)^{1/2},
\end{eqnarray}
where $\alpha^{\mathrm{(B)}}_4 = 2.59\times 10^{-13}$ cm$^3$ s$^{-1}$ is the case B recombination coefficient at $10^4$ K, $T_{\rm II} = 10^4$ is the typical H~\textsc{ii} region temperature, $f_{\rm trap} = 2$ is a trapping factor that accounts for stellar wind and trapped infrared radiation pressure, $\psi = 3.2$  is the mean photon energy in Rydberg for a fully sampled IMF at zero age\red{\footnote{\red{This value of $\psi$ is taken from \citet{murray10b}, and is based on a \citet{chabrier05a} IMF and a compilation of empirically-measured ionizing luminosities for young stars. However, alternative IMFs and methods of computing the ionizing luminosity give results that agree to tens of percent.}}}, and $\mu = 1.33$ is the mean mass per hydrogen nucleus for gas \red{with the usual helium abundance $x_{\mathrm{He}} = 0.1$}. The solution includes the effects of both radiation and gas pressure in driving the expansion. Once the radius is known, the density near the H~\textsc{ii} region center (the parameter required by \texttt{cloudy}) can be computed from the usual ionization balance argument,
\begin{equation}
n_{\rm II} = \left(\frac{3 Q(\mathrm{H}^0)}{4.4\pi\alpha^{\mathrm{(B)}}_4 r_{\rm II}^3}\right)^{1/2}.
\end{equation}
Note that the factor of $4.4$ in the denominator accounts for the extra free electrons provided by helium, assuming it is singly ionized. Also note that this will not give the correct density during the brief radiation pressure-dominated phase early on in the evolution, but that this period is very brief (though the effects of the extra boost provided by radiation pressure can last much longer), and no simple analytic approximation for this density is available \citep{krumholz09d}. \red{Also note that, although \cloudyslug~presently only supports this parameterization of H~\textsc{ii} region radius and density versus time, the code is simply a python script. It would therefore be extremely straightforward for a user who prefers a different parameterization to alter the script to supply it.}

In cluster mode, \cloudyslug\ uses the approximation described above to compute the density of the ionized gas in the vicinity of each star cluster, which is then passed as an input to \texttt{cloudy} along with the star cluster's spectrum. Note that computations in cluster mode are much more expensive than those in integrated mode, since the latter requires only a single call to \texttt{cloudy} per time step, while the former requires one per cluster. To ease the computational requirements slightly, in cluster mode one can set a threshold ionizing luminosity below which the contribution to the total nebular spectrum is assumed to be negligible, and is not computed.

\subsection{\bp: Bayesian Inference from Stellar Photometry}
\label{ssec:bp}

\subsubsection{Description of the Method}

Simulating stochastic stellar populations is useful, but to fully exploit this capability we must tackle the inverse problem: given an observed set of stellar photometry, what should we infer about the physical properties of the underlying stellar population in the regime where the mapping between physical and photometric properties is non-deterministic? A number of authors have presented numerical methods to tackle problems of this sort, mostly in the context of determining the properties of star clusters with stochastically-sampled IMFs \citep{popescu09a, popescu10a, popescu10b, fouesneau10a, fouesneau12a, popescu12a, asad12a, anders13a, de-meulenaer13a, de-meulenaer14a, de-meulenaer15a}, and in some cases in the context of determining SFRs from photometry \citep{da-silva14b}. Our method here is a generalization of that developed in \citet{da-silva14b}, and has a number of advantages as compared to earlier methods, both in terms of its computational practicality and its generality.

Consider stellar systems characterized by a set of $N$ physical parameters $\vecx = (x_1, x_2, \ldots x_N)$; in the example of star clusters below we will have $N=3$, with $x_1$, $x_2$, and $x_3$ representing the logarithm of the mass, the logarithm of the age, and the extinction, while for galaxies forming stars at a continuous rate we will have $N=1$, with $x_1$ representing the logarithm of the SFR. The light output of these systems is known in $M$ photometric bands; let $\vecy = (y_1, y_2, \ldots y_M)$ be a set of photometric values, for example magnitudes in some set of filters. Suppose that we observe a stellar population in these bands and measure a set of photometric values $\vecyo$, with some errors $\vecs = (\sigma_{y_1}, \sigma_{y_2}, \ldots, \sigma_{y_M})$, which for simplicity we will assume are Gaussian. We wish to infer the posterior probability distribution for the physical parameters given the observed photometry and photometric errors, $p(\vecx \mid \vecyo; \vecs)$. 

Following \citet{da-silva14b}, we compute the posterior probability via implied conditional regression coupled to kernel density estimation. Let the joint PDF of physical and photometric values for the population of all the stellar populations under consideration be $p(\vecx, \vecy)$. We can write the posterior probability distribution we seek as
\begin{equation}
\label{eq:decomposition}
p(\vecx \mid \vecy) \propto \frac{p(\vecx, \vecy)}{p(\vecy)},
\end{equation}
where $p(\vecy)$ is the distribution of the photometric variables alone, i.e.,
\begin{equation}
\label{eq:py}
p(\vecy) \propto \int p(\vecx, \vecy) \, d\vecx.
\end{equation}
If we have an exact set of photometric measurements $\vecyo$, with no errors, then the denominator in equation (\ref{eq:decomposition}) is simply a constant that will normalize out, and the posterior probability distribution we seek is distributed simply as
\begin{equation}
\label{eq:noerror}
p(\vecx \mid \vecyo) \propto  p(\vecx, \vecyo).
\end{equation}
In this case, the problem of computing $p(\vecx \mid \vecyo)$ reduces to that of computing the joint physical-photometric PDF $p(\vecx, \vecy)$ at any given set of observed photometric values $\vecyo$.

For the more general case where we do have errors, we first note that the posterior probability distribution for the true photometric value $\vecy$ is given by the prior probability distribution for photometric values multiplied by the likelihood function associated with our measurements. The prior PDF of photometric values is simply $p(\vecy)$ as given by equation (\ref{eq:py}), and for a central observed value of $\vecyo$ with errors $\vecs$, the likelihood function is simply a Gaussian. Thus the PDF of $\vecy$ given our observations is
\begin{equation}
\label{eq:errors}
p(\vecy \mid \vecyo) \propto p(\vecy) G(\vecy-\vecyo, \vecs)
\end{equation}
where
\begin{equation}
G(\vecy, \vecs) \propto \exp\left[-\left(\frac{y_1^2}{2\sigma_{y_1}^2} + \frac{y_2^2}{2\sigma_{y_2}^2} + \cdots + \frac{y_M^2}{2\sigma_{y_M}^2}\right)\right]
\end{equation}
is the usual multi-dimensional Gaussian function. The posterior probability for the physical parameters is then simply the convolution of equations (\ref{eq:decomposition}) and (\ref{eq:errors}), i.e.,
\begin{eqnarray}
p(\vecx \mid \vecyo) & \propto & \int p(\vecx \mid \vecy) \, p(\vecy \mid \vecyo) \, d\vecy
\nonumber \\
& \propto &
\int p(\vecx, \vecy) \, G(\vecy-\vecyo, \vecs) \, d\vecy.
\label{eq:pdffinal}
\end{eqnarray}
Note that we recover the case without errors, equation (\ref{eq:noerror}), in the limit where $\vecs \rightarrow 0$, because in that case the Gaussian $G(\vecy-\vecyo, \vecs) \rightarrow \delta(\vecy-\vecyo)$.

We have therefore reduced the problem of computing $p(\vecx \mid \vecyo)$ to that of computing $p(\vecx, \vecy)$, the joint PDF of physical and photometric parameters, for our set of stellar populations. To perform this calculation, we use \slug\ to create a large library of models for the type of stellar population in question. We draw the physical properties of the stellar populations (e.g., star cluster mass, age, and extinction) in our library from a distribution $p_{\mathrm{lib}}(\vecx)$, and for each stellar population in the library we have a set of physical and photometric parameters $\vecx_i$ and $\vecy_i$. We estimate $p(\vecx, \vecy)$ using a kernel density estimation technique. Specifically, we approximate this PDF as
\begin{equation}
\label{eq:KDE}
p(\vecx, \vecy) \propto \sum_i w_i K(\vecx - \vecx_i, \vecy - \vecy_i; \vech),
\end{equation}
where $w_i$ is the weight we assign to sample $i$, $K(\vecx; \vech)$ is our kernel function, $\vech = (h_{x_1}, h_{x_2}, \ldots, h_{x_N}, h_{y_1}, h_{y_2}, \ldots, h_{y_M})$ is our bandwidth parameter (see below), and the sum runs over every simulation in our library. We assign weights to ensure that the distribution of physical parameters $\vecx$ matches whatever prior probability distributions we wish to assign for them. If we let $p_{\mathrm{prior}}(\vecx)$ represent our priors, then this is
\begin{equation}
\label{eq:weight}
w_i = \frac{p_{\mathrm{prior}}(\vecx_i)}{p_{\mathrm{lib}}(\vecx_i)}.
\end{equation}
Note that, although we are free to choose $p_{\mathrm{lib}}(\vecx) = p_{\mathrm{prior}}(\vecx)$ and thus weight all samples equally, it is often advantageous for numerical or astrophysical reasons not to do so, because then we can distribute our sample points in a way that is chosen to maximize our knowledge of the shape of $p(\vecx, \vecy)$ with the fewest possible realizations. \red{As a practical example, we know that photometric outputs will fluctuate more for galaxies with low star formation rates than for high ones, so the library we use for \sfrslug~(see below) is weighted to contain more realizations at low than high SFR.}

We choose to use a Gaussian kernel function, $K(\vecx; \vech) = G(\vecx, \vech)$, because this presents significant computational advantages. Specifically, with this choice, equation (\ref{eq:pdffinal}) becomes
\begin{eqnarray}
\lefteqn{p(\vecx \mid \vecyo) 
} \nonumber \\
& \propto & \sum_i w_i G(\vecx-\vecx_i, \vech_x)
\nonumber \\
& & \quad 
\int G(\vecy - \vecyo, \vecs) \,G(\vecy - \vecy_i, \vech_y) \,d\vecy 
\label{eq:gaussconvol1}
\\
& \propto & \sum_i w_i G(\vecx-\vecx_i, \vech_x)
\nonumber \\
& & \quad G(\vecyo - \vecy_i, \sqrt{\vecs^2+\vech_y^2}) 
\label{eq:gaussconvol2}
\\
& \equiv & \sum_i w_i G((\vecx-\vecx_i, \vecyo-\vecy_i), \vech')
\label{eq:pdf}
\end{eqnarray}
where $\vech_x = (h_{x_1}, h_{x_2}, \ldots, h_{x_N})$ is the bandwidth in the physical dimensions,
$\vech_y = (h_{y_1}, h_{y_2}, \ldots, h_{y_M})$ is the bandwidth in the photometric dimensions, and the quadrature sum $\sqrt{\vech_y^2+\vecs^2}$ is understood to be computed independently over every index in $\vecs$ and $\vech_y$. The new quantity we have introduced,
\begin{equation}
\vech' = (\vech_x, \sqrt{\vech_y^2+\vecs^2}),
\end{equation}
is simply a modified bandwidth in which the bandwidth in the photometric dimensions has been broadened by adding the photometric errors in quadrature sum with the bandwidths in the corresponding dimensions. Note that, in going from equation (\ref{eq:gaussconvol1}) to equation (\ref{eq:gaussconvol2}) we have invoked the result that the convolution of two Gaussians is another Gaussian, whose width is the quadrature sum of the widths of the two input Gaussians, and whose center is located at the difference between the two input centers.

As an adjunct to this result, we can also immediately write down the marginal probabilities for each of the physical parameters in precisely the same form. The marginal posterior probability distribution for $x_1$ is simply
\begin{eqnarray}
p(x_1 \mid \vecyo) & \propto & \int p(\vecx \mid \vecyo) \, dx_2 \, dx_3 \ldots dx_N \\
& \propto & \sum_i w_i G(x_1 - x_{1,i}, h_1) 
\nonumber \\
& & \quad
G(\vecyo-\vecy_i, \sqrt{\vecs^2+\vech_y^2}),
\label{eq:mpdf}
\end{eqnarray}
and similarly for all other physical variables. This expression also immediately generalizes to the case of joint marginal probabilities of the physical variables. We have therefore succeeded in writing down the posterior probability distribution for all the physical properties, and the marginal posterior distributions of any of them individually or in combination, via a kernel density estimation identical to that given by equation (\ref{eq:KDE}). \red{One advantage of equations (\ref{eq:pdf}) and (\ref{eq:mpdf}) is that they are easy to evaluate quickly using standard numerical methods. Details on our software implementation are given in Appendix \ref{app:software}.}

\subsubsection{\red{Error Estimation and Properties of the Library}}
\label{sssec:errorest}

\red{
An important question for \bp, or any similar method, is how large a library must be in order to yield reliable results. Of course this depends on what quantity is being estimated and on the shape of the underlying distribution. Full, rigorous, error analysis is best accomplished by bootstrap resampling. However, we can, without resorting to such a computationally-intensive procedure, provide a useful rule of thumb for how large a library must be so that the measurements rather than the library are the dominant source of uncertainty in the resulting derived properties.}

\red{Consider a library of $n$ simulations, from which we wish to measure the value $X_q$ that delineates a certain quantile (i.e., the percentile $p = 100q$) of the underlying distribution from which the samples are drawn. Let $x_i$ for $i=1 \ldots n$ be the samples in our library, ordered from smallest to largest. Our central estimate for the value of $X_q$ that delineates quintile $q$, will naturally be $x_{qn}$. (Throughout this argument we will assume that $n$ is large enough that we need not worry about the fact that ranks in the list, such as $qn$, will not exactly be integers. Extending this argument to include the necessary continuity correction adds mathematical complication but does not change the basic result.) For example, we may have a library of $10^6$ simulations of galaxies at a particular SFR (or in a particular interval of SFRs), and wish to know what ionizing luminosity corresponds to the 95th percentile at that SFR. Our estimate for the 95th percentile ionizing luminosity will simply be the ionizing luminosity of the 950,000th simulation in the library.}

\red{
To place confidence intervals around this estimate, note that the probability that a randomly chosen member of the population will have a value $x < x_q$ is $q$, and conversely the probability that $x>x_q$ is $1-q$. Thus in our library of $n$ simulations, the probability that we have exactly $k$ samples for which $x<x_q$ is given by the binomial distribution,
\begin{equation}
\mathrm{Pr}(k) = \frac{n!}{(n-k)!} q^k (1-q)^{n-k}.
\end{equation}
When $n\gg 1$, $k\gg 1$, and $n-k \gg 1$, the binomial distribution approaches a Gaussian distribution of central value $qn$ and standard deviation $\sqrt{q(1-q)n}$:
\begin{equation}
\mathrm{Pr}(k) \approx \frac{1}{\sqrt{2\pi q(1-q)n}} \exp\left[-\frac{(k-qn)^2}{2 q(1-q)n}\right].
\end{equation}
That is, the position $n$ in our library of simulations such that $x_i<x_q$ for all $i<n$, and $x_i>x_q$ for all $i>n$, is distributed approximately as a Gaussian, centered at $qn$, with dispersion $\sqrt{q(1-q)n}$. This makes it easy to compute confidence intervals on $x_q$, because it reduces the problem that of computing confidence intervals on a Gaussian. Specifically, if we wish to compute the central value and confidence interval $c$ (e.g., $c=0.68$ is the 68\% confidence interval) for the rank $r$ corresponding to percentile $q$, this is
\begin{equation}
r \approx qn \pm \sqrt{2q(1-q)n}\,\mathrm{erf}^{-1}(c).
\end{equation}
The corresponding central value for $X_q$ (as opposed to its rank in the list) is $x_{qn}$, and the confidence interval is
\begin{equation}
\left(x_{qn - \sqrt{2q(1-q)n}\,\mathrm{erf}^{-1}(c)}, x_{qn+\sqrt{2q(1-q)n}\,\mathrm{erf}^{-1}(c)}\right]).
\end{equation}
In our example of determining the 95th percentile from a library of $10^6$ simulations, our central estimate of this value is the $x_{950,000}$, the 950,000th simulation in the library, and the 90\% confidence interval is 359 ranks on either side of this. That is, our 90\% confidence interval extends from $x_{949,641}$ to $x_{950,359}$.
}

\red{
This method may be used to define confidence intervals on any quantile derived from a library of \slug~simulations. Of course this does not specifically give confidence intervals on the results derived when such a library is used as the basis of a Bayesian inference from \bp. However, this analysis can still provide a useful constraint: the error on the values derived from \bp~cannot be less than the errors on the underlying photometric distribution we have derived from the library. Thus if we have a photometric measurement that lies at the $q$th quantile of the library we are using for kernel density estimation in \bp, the effective uncertainty our kernel density estimation procedure provides is at a minimum given by the uncertainty in the quantile value $x_q$ calculated via the procedure above. If this uncertainty is larger than the photometric uncertainties in the measurement, then the resolution of the library rather than the accuracy of the measurement will be the dominant source of error.
}

\subsection{\texttt{sfr\_slug} and \texttt{cluster\_slug}: Bayesian Star Formation Rates and Star Cluster Properties}
\label{ssec:sfr_cluster_slug}

The \slug\ package ships with two Bayesian inference modules based on \bp. \Sfrslug, first described in \citet{da-silva14b} and modified slightly to use the improved computational technique described in the previous section, is a module that infers the posterior probability distribution of the SFR given an observed flux in H$\alpha$, GALEX FUV, or bolometric luminosity, or any combination of the three. The library of models on which it operates (which is included in the \slug\ package) consists of approximately $1.8\times 10^6$ galaxies with constant SFRs sampling a range from $10^{-8} - 10^{0.5}$ $M_\odot$ yr$^{-1}$, with no extinction; the default bandwidth is 0.1 dex. We refer readers for \citet{da-silva14b} for a further description of the model library. \Sfrslug\ can download the default library automatically, and it is also available as a standalone download from \url{http://www.slugsps.com/data}.

The \clusterslug\ package performs a similar task for inferring the mass, age, and extinction of star clusters. The \slug\ models that \clusterslug\ uses consist of \red{libraries of $10^7$} simple stellar populations with masses in the range $\log (M/M_\odot) = 2 - 8$, ages $\log (T/\mbox{yr}) = \red{5 - \log T_{\mathrm{max}}}$, and extinctions $A_V = 0 - 3$ mag. \red{The maximum age $T_{\mathrm{max}}$ is either 1 Gyr or 15 Gyr, depending on the choice of tracks (see below). The data are sampled uniformly in $A_V$. In mass the library sampling is $dN/dM \propto M^{-1}$ for masses up to $10^4$ $M_\odot$, and as $dN/dM \propto M^{-2}$ at higher masses; similarly, the library age sampling is $dN/dT \propto T^{-1}$ for $T < 10^8$ yr, and as $dN/dT \propto T^{-2}$ at older ages. The motivation for this sampling is that it puts more computational effort at younger ages and low masses, where stochastic variation is greatest. The libraries all use a \citet{chabrier05a} IMF, and include nebular emission computed using $\phi=0.73$ and an ionization parameter $\log\,\mathcal{U} = -3$ (see Appendix \ref{app:nebuladust}). Libraries are available using either Padova or Geneva tracks, with the former having a maximum age of 15 Gyr and the latter a maximum age of 1 Gyr. The Geneva tracks are available using either Milky Way or ``starburst" extinction curves} (see \red{Appendix \ref{app:nebuladust}}). The default bandwidth is 0.1 dex in mass and age, 0.1 mag in extinction, and 0.25 mag (corresponding to 0.1 dex in luminosity) in photometry. For each model, photometry is computed for a large range of filters, listed in \autoref{tab:cs_filters}. As with \sfrslug, \clusterslug\ can automatically download the default library, and the data are also available as a standalone download from \url{http://www.slugsps.com/data}. Full spectra for the library, allowing the addition of further filters as needed, are also available upon request; they are not provided for web download due to the large file sizes involved.

\begin{table*}
\centering
\caption{Filters in the \texttt{cluster\_slug} library}
\label{tab:cs_filters}
\begin{tabular}{ll}
\hline
Type & Filter Name \\ \hline\hline
\textit{HST WFC3 UVIS} wide & F225W, F275W, F336W, F360W, F438W, F475W, F555W, F606W, F775W, F814W \\
\textit{HST WFC3 UVIS} medium/narrow & F547M, F657N, F658N \\
\textit{HST WFC3 IR} & F098M, F105W, F110W, F125W, F140W, F160W \\
\textit{HST ACS} wide & F435W, F475W, F555W, F606W, F625W, F775W, F814W \\
\textit{HST ACS} medium/narrow & F550M, F658N, F660N \\
\textit{HST ACS HRC} & F330W \\
\textit{HST ACS SBC} & F125LP, F140LP, F150LP F165LP \\
\textit{Spitzer} IRAC & 3.6, 4.5, 5.8, 8.0 \\
\textit{GALEX} & FUV, NUV \\
Johnson-Cousins & U, B, V, R, I \\
SDSS & u, g, r, i, z \\
2MASS & J, H, Ks\\
\hline
\end{tabular}
\end{table*}

\section{Sample Applications}
\label{sec:applications}

In this section, we present a suite of test problems with the goal of illustrating the 
different capabilities of \slug, highlighting the effects of  key parameters on \slug's simulations, 
and validating the code. Unless otherwise stated, all computations use the Geneva (2013) non-rotating 
stellar tracks and stellar atmospheres following the \verb=starburst99= implementation.

\begin{figure}
\includegraphics[width=\columnwidth]{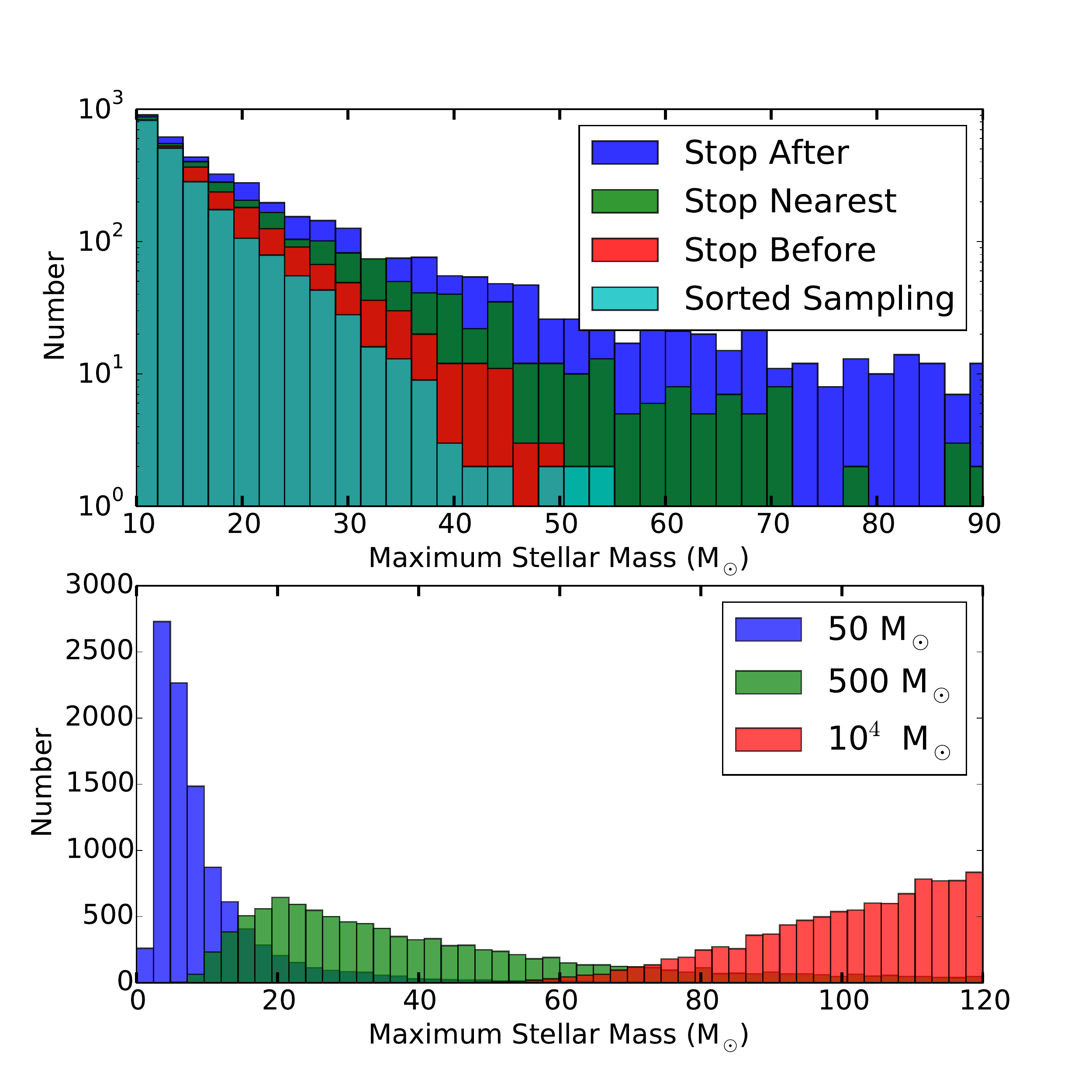}
\caption{
\label{fig:sampltec}
Histograms of the mass of the most massive stars in sets of \red{$10^4$ ``cluster'' simulations}, which were
performed with different sampling techniques (top panel) and different target cluster masses  
(bottom panel). In the top panel, four different stop criteria are adopted to sample a 
\citet{kroupa02c} IMF for clusters of 50 M$_\odot$. In the bottom panel, the default 
{\tt STOP\_NEAREST} condition is chosen to sample from a \citet{kroupa02c} IMF in clusters 
of three different masses.}
\end{figure}

\subsection{Sampling Techniques}

As discussed above and in the literature \red{\citep[e.g.,][]{weidner06a,haas10a,da-silva12a, cervino13b, popescu14a}}, the choice of 
sampling technique used to generate stellar masses can have significant effects on the \red{distribution of stellar masses and the} final light output, 
even when the underlying distribution being sampled is held fixed. \red{This can have profound astrophysical implications. Variations in sampling technique even for a fixed IMF can produce systematic variations in galaxy colors \citep[e.g.,][]{haas10a}, nucleosynthetic element yields \citep[e.g.,][]{koppen07a, haas10a}, and observational tracers of the star formation rate \citep[e.g.,][]{weidner04b, pflamm-altenburg07b}. It is therefore of interest to explore how sampling techniques influence various aspects of stellar populations. We do so} as a first demonstration of \slug's capabilities. Here we consider the problem of populating clusters with stars from an IMF,
but our discussion also applies to the case of ``galaxy'' simulations.
Specifically, we run four sets of \red{$10^4$ ``cluster'' simulations} with a target mass of 50 M$_\odot$
by sampling a \citet{kroupa02c} 
IMF with the {\tt STOP\_NEAREST}, {\tt STOP\_AFTER}, {\tt STOP\_BEFORE}, 
{\tt SORTED\_SAMPLING} conditions \red{(see Appendix \ref{app:pdfs})}. In the following, we analyse a single timestep at $10^6$ yr.

By default, \slug\ adopts the {\tt STOP\_NEAREST} condition, according to which the final draw
from the IMF is added to the cluster only if the inclusion of this last star minimises 
the absolute error between the target and the achieved cluster mass. In this case, the cluster mass
sometimes exceeds and sometimes falls short of the desired mass. 
The {\tt STOP\_AFTER} condition, instead, always includes the 
final draw from the IMF. Thus, with this choice, \slug\ simulations 
produce clusters with masses that are always in excess of the target mass. 
The opposite behaviour is obviously recovered by the {\tt STOP\_BEFORE} condition, 
in which the final draw is always rejected. Finally, for {\tt SORTED\_SAMPLING} condition, 
the final cluster mass depends on the details of the chosen IMF.

Besides this manifest effect of the sampling techniques on the achieved cluster masses, 
the choice of sampling produces a drastic effect on the distribution of stellar masses, 
even for a fixed IMF. This is shown in the top panel of \autoref{fig:sampltec}, 
where we display histograms for the mass of the most massive stars within these simulated clusters. 
One can see that, compared to the default {\tt STOP\_NEAREST} condition, 
the {\tt STOP\_AFTER} condition
results in more massive stars being included in the simulated clusters. Conversely, 
the {\tt STOP\_BEFORE} and the {\tt SORTED\_SAMPLING} undersample the massive end of the 
IMF. Such a different stellar mass distribution has direct implications for the photometric 
properties of the simulated stellar populations, especially for wavelengths that are sensitive 
to the presence of most massive stars. \red{Similar results have previously been obtained by
other authors, including \citet{haas10a} and \citet{cervino13b}, but prior to \slug~no publicly-available
code has been able to tackle this problem.}

The observed behaviour on the stellar mass distribution for a particular choice of sampling 
stems from a generic mass constraint: clusters
cannot be filled with stars that are more massive than the target cluster itself. 
\Slug\ does not enforce this condition strictly, allowing for realizations in which 
stars more massive than the target cluster mass are included. However, 
different choices of the sampling technique result in 
a different degree with which \slug\ enforces this mass constraint.
To further illustrate the relevance of this effect,
we run three additional ``cluster'' simulations assuming again 
a \citet{kroupa02c} IMF and the default {\tt STOP\_NEAREST} condition. Each simulation is composed 
\red{by $10^4$ trials}, but with a target cluster mass of $M_{\rm cl,t} = 50$, 500, and $10^4$ M$_\odot$. 
The bottom panel of \autoref{fig:sampltec} shows again the mass distribution of the most massive 
star in each cluster. As expected, while massive stars cannot be found  in low mass clusters, 
the probability of finding at least a star as massive as the IMF upper limit 
increases with the cluster mass. At the limit of very large masses, a nearly fully-sampled 
IMF is recovered as the mass constraint becomes almost irrelevant. \red{Again, similar results
have previously been obtained by \citet{cervino13a}.}

\begin{figure*}
\includegraphics[width=0.75\textwidth]{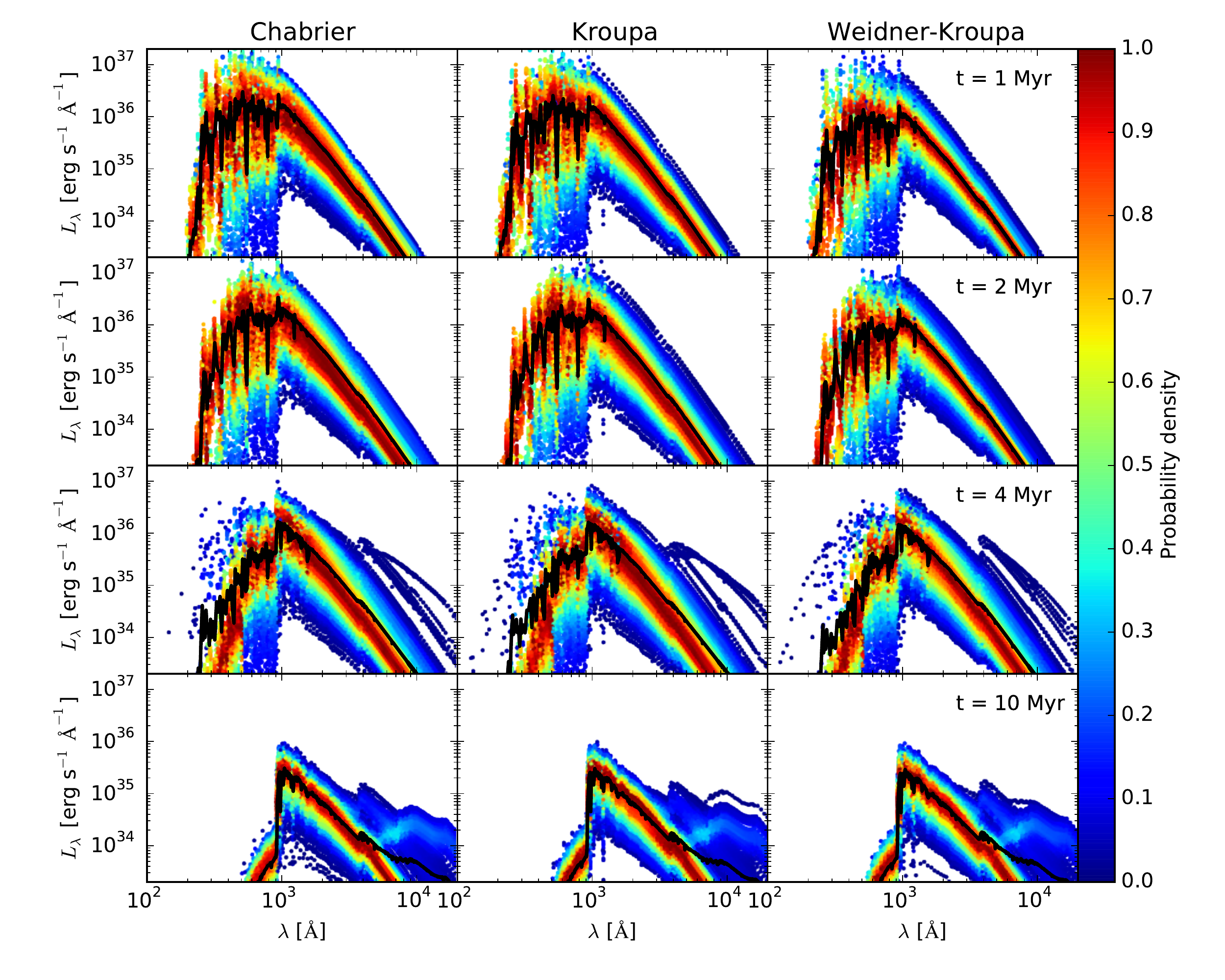}
\caption{\label{fig:imfvary1}
Spectra of simple stellar populations at ages of $1, 2, 4$ and $10$ Myr (top to bottom rows) for the IMFs of \citet[left column]{chabrier05a}, \citet[middle column]{kroupa02c}, and \citet[left column]{weidner06a}. In each panel, thick black lines indicate the mean, and \red{points show the locations of individual simulations (10\% of the simulations, selected at random), with the color of the point indicating the probability density for the monochromatic luminosity at each wavelength (see color bar) Probability densities are evaluated independently at each monochromatic luminosity, and are normalized to have a maximum of unity at each wavelength.}.
}
\end{figure*}

\subsection{Stochastic Spectra and Photometry for Varying IMFs}

Together with the adopted sampling technique, the choice of the IMF is an important 
parameter that shapes \slug\ simulations. We therefore continue the demonstration of \slug's 
capabilities by computing spectra and photometry for simple stellar populations with total masses 
of 500 $M_\odot$ for three different IMFs. We create 1000 realizations each of such a population at times from $1-10$ Myr in intervals of 1 Myr, using the IMFs of \citet{chabrier05a}, \citet{kroupa02c}, and \citet{weidner06a}. The former two IMFs use \verb=STOP_NEAREST= sampling, while the latter uses \verb=SORTED_SAMPLING=, and is otherwise identical to the \citet{kroupa02c} IMF. 

\begin{figure}
\includegraphics[width=\columnwidth]{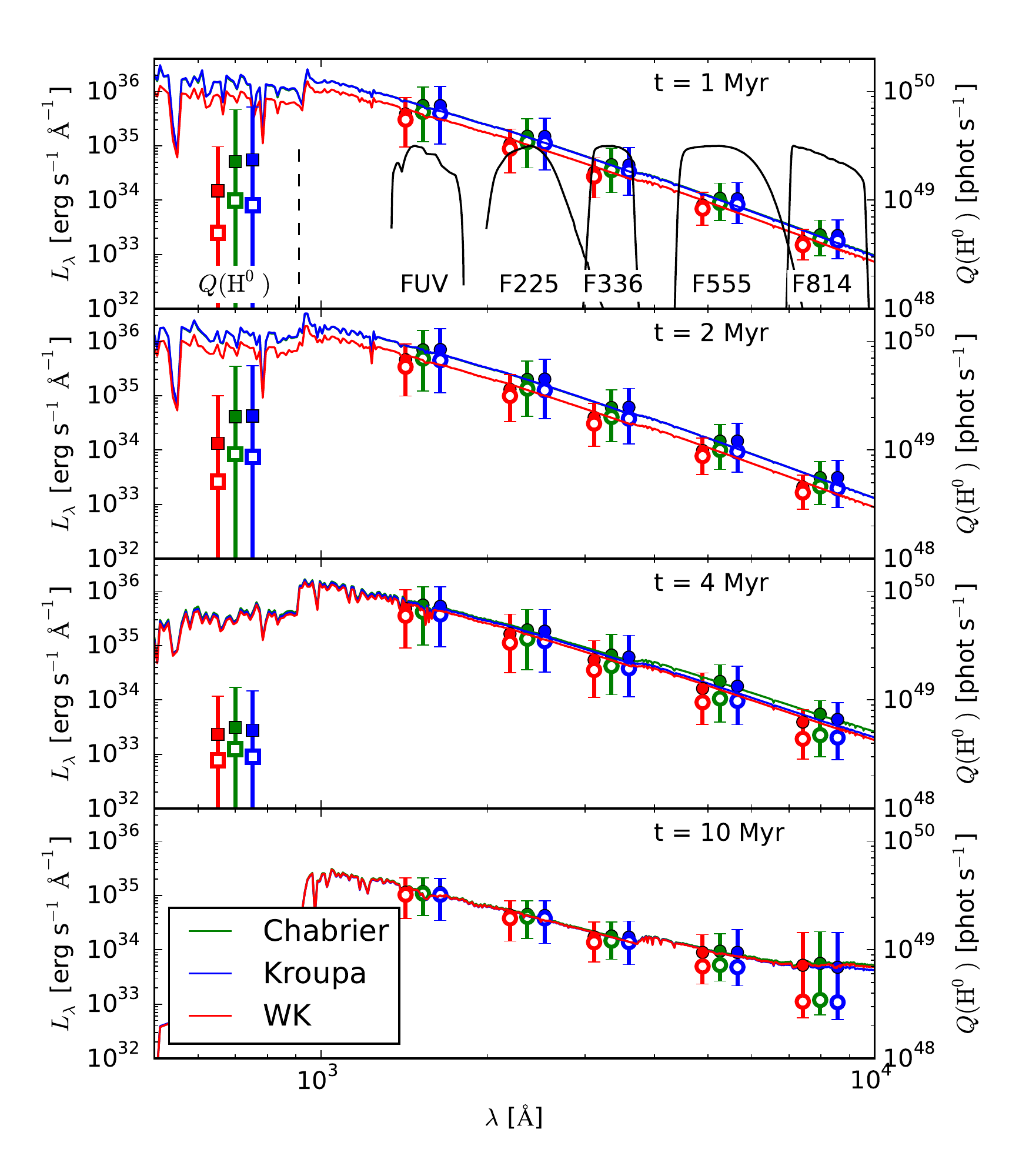}
\caption{\label{fig:imfvary2}
Photometry of simple stellar populations at the same times as shown in \autoref{fig:imfvary1}. In each panel, the thin lines show the same mean spectra plotted in \autoref{fig:imfvary1}. Circles with error bars show the specific luminosity $L_\lambda$ measured in each of the indicated filters: \textit{GALEX} FUV, and \textit{HST UVIS} F225W, F336W, F555W, and F814W. The abcissae of the green points are placed at the effective wavelength of each filter, with the red and blue offset to either side for clarity. The labeled black curves in the top panel show the filter response functions for each filter. For these points, filled circles indicate the mean value, open circles indicate the median value, and error bars indicate the range from the $10-90$th percentile. The leftmost, square points with error bars show the comparable mean, median, and range for the ionizing luminosity (scale on the right axis).
}
\end{figure}

\begin{figure}
\includegraphics[width=\columnwidth]{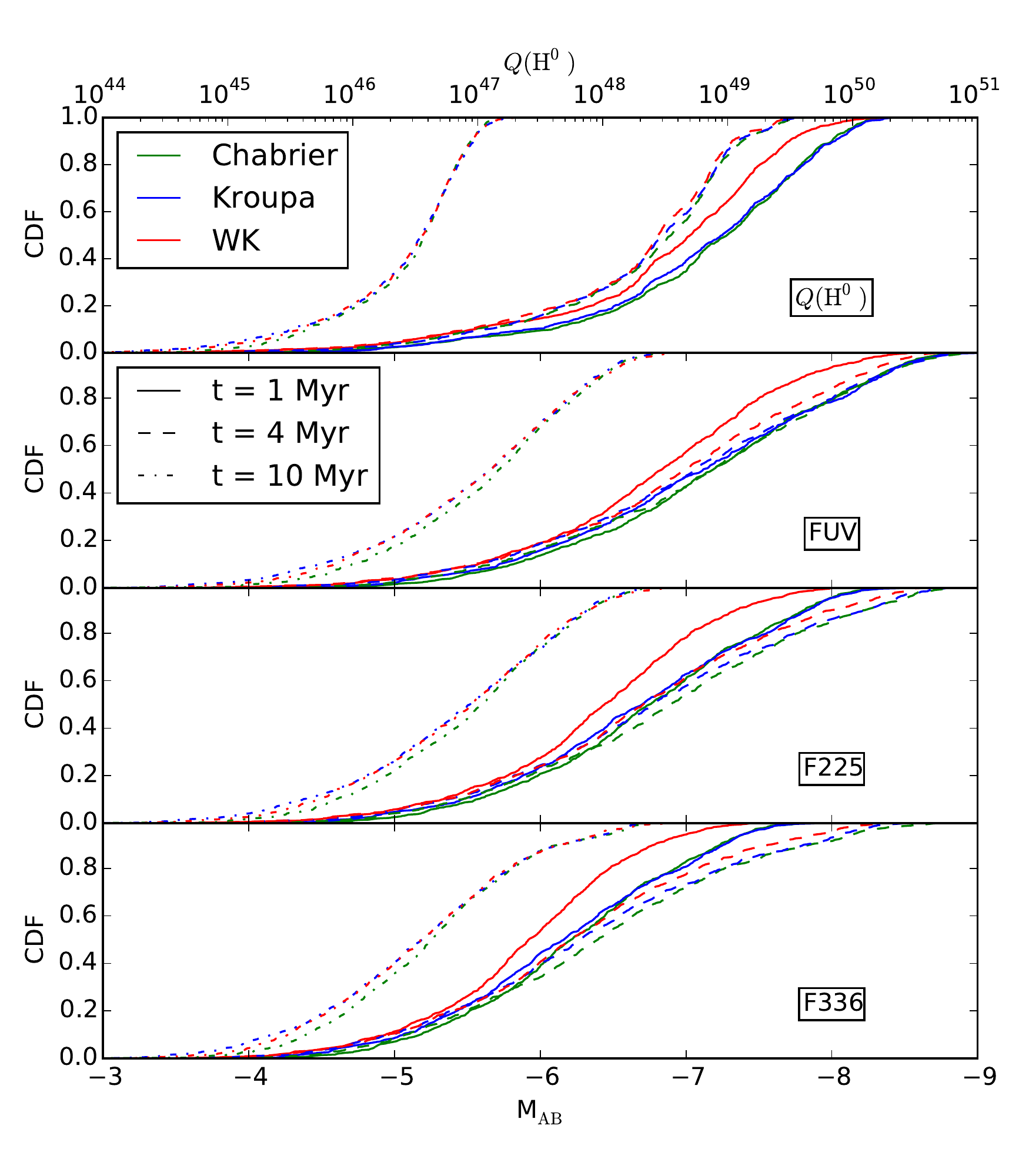}
\caption{\label{fig:imfvary3}
Cumulative distribution functions for ionizing luminosity (top panel) and in the \textit{GALEX} FUV and \textit{HST UVIS} F225W and F336W filters (bottom three panels). In each panel, the $x$ axis shows either the ionizing luminosity (for the top panel) or the absolute AB magnitude (bottom 3 panels) of a $500$ $M_\odot$ simple stellar population drawn from a \citet[green]{chabrier05a}, \citet[blue]{kroupa02c}, or \citet[red]{weidner06a} IMF at ages of 1 Myr (solid), 4 Myr (dashed), and 10 Myr (dot-dashed).
}
\end{figure}

Figures \ref{fig:imfvary1} -- \ref{fig:imfvary3} show the distributions of spectra and photometry that result from these simulations. \red{Stochastic variations in photometry have been studied previously by a number of authors, going back to \citet{chiosi88a}, but to our knowledge no previous authors have investigated similar variations in spectra.} The plots also demonstrate \slug's ability to evaluate the full probability distribution for both spectra and photometric filters, and reveal interesting phenomena that would not be accessible to a non-stochastic SPS code. In particular, \autoref{fig:imfvary1} shows that the mean specific luminosity can be orders of magnitude \red{larger than the mode}. The divergence is greatest at wavelengths of a few hundred \AA\ at ages $\sim 2-4$ Myr, and wavelengths longer than $\sim 5000$ \AA\ at $10$ Myr. Indeed, at 4 Myr, it is noteworthy that the mean spectrum is actually outside the $10-90$th percentile range. In this particular example, the range of wavelengths at 4 Myr where the mean spectrum is outside the $10-90$th percentile corresponds to energies of $\sim 3$ Ryd. For a stellar population 4 Myr old, these photons are produced only by WR stars\red{, and most prolifically by extremely hot WC stars. It turns out that a WC star is present $\sim 5\%$ of the time, but these cases are so much more luminous than when a WC star is not present that they are sufficient to drag the mean upward above the highest luminosities that occur in the $\sim 95\%$ of cases when no WC star is present.}

A similar phenomenon is visible in the photometry shown by \autoref{fig:imfvary2}. In most of the filters the $10-90$th percentile range is an order of magnitude wide, and for the ionizing luminosity and the \textit{HST UVIS} F814W at 10 Myr the spread is more than two orders of magnitude, with significant offsets between mean and median indicating a highly asymmetric distribution. \autoref{fig:imfvary3}, which shows the full distributions for several of the filters, confirms this impression: at early times the cumulative distribution functions are extremely broad, and the ionizing luminosity in particular shows a broad distribution at low $Q(\mathrm{H}^0)$ and then a small number of simulations with large $Q(\mathrm{H}^0)$. \red{Note that the percentile values are generally very well-determined by our set of 1000 simulations. Using the method described in \autoref{sssec:errorest} to compute the errors on the percentiles, we find that the 68\% confidence interval on the 10th, 50th, and 90th percentile values is less than 0.1 dex wide at almost all times, filters, and IMFs. The only exception is at 1 Myr, where the 68\% confidence interval on the 10th percentile of ionizing luminosity is $\sim 0.2-0.3$ dex wide.}

The figures also illustrate \slug's ability to capture the ``IGIMF effect'' \citep{weidner06a} whereby sorted sampling produces a systematic bias toward lower luminosities at short wavelengths and early times. Both the spectra and photometric values for the \citet{weidner06a} IMF are systematically suppressed relative to the IMFs that use a sampling method that is less biased against high mass stars (cf. \autoref{fig:sampltec}). 

\begin{figure}
\includegraphics[width=\columnwidth]{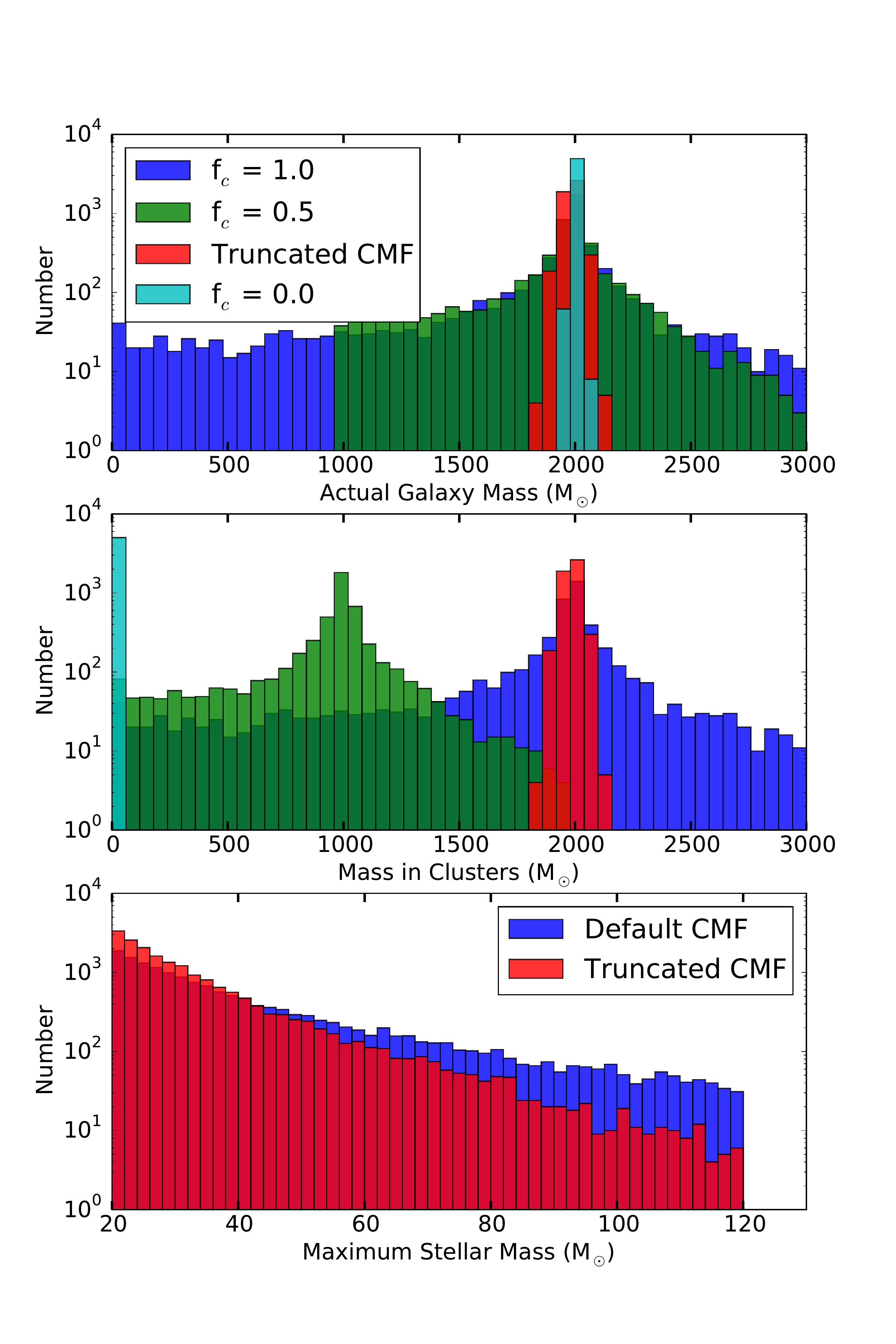}
\caption{\label{fig:clusterprop}
\red{Histograms of the actual galaxy mass (top), total stellar 
mass in clusters (middle), and \red{mass of the most massive star} formed 
in clusters (bottom) for four different \slug\ simulations with 
5000 realizations each}. The three simulations labeled as $f_c$ consist of 
the default \slug\ CMF and different choices of the fraction of stars formed within clusters. 
The bottom panel compares instead two simulations with $f_c=1$, but for two different 
choices of CMF (\slug's default and a CMF truncated at 100 M$_\odot$).}
\end{figure}

\subsection{Cluster Fraction and Cluster Mass Function}
\label{ssec:clustfract}

The first two examples have focused on simple stellar populations, namely collections of 
stars that are formed at coeval times to compose a ``cluster'' simulation. Our third
example highlights \slug's ability to simulate composite stellar populations. Due to
\slug's treatment of stellar clustering and stochasticity, simulations of such
populations in \slug\ differ substantially from those in non-stochastic SPS codes. Unlike in a
conventional SPS code, the outcome of a \slug~simulation is sensitive to both the
CMF and the fraction $f_{\rm c}$ of stars formed in clusters. \red{\slug\ can therefore
be applied to a wide variety of problems in which the fraction of stars formed 
within clusters or in the field becomes a critical parameter \citep[e.g.][]{fumagalli11a}.}

\red{The relevance of clusters in \slug\ simulations arises from two channels}.
First, because stars form in clusters of finite size, and the interval between cluster formation events is not necessarily large compared to the lifetimes of individual massive stars, clustered star formation produces significantly more variability in the number of massive stars present at any given time than non-clustered star formation. We defer a discussion of this effect to the following section. Here, we instead focus on a second channel by which the CMF and $f_{\rm c}$ influence the photometric output, which arises due to
the method by which \slug\ implements mass constraints.
As discussed above, a realization of the IMF in a given cluster simulation is 
the result of the mass constraint imposed by the mass of the cluster, drawn from a CMF, 
and by the sampling technique chosen to approximate the target mass. Within ``galaxy'' simulations,
a second mass constraint is imposed on the simulations as the SFH provides 
an implicit target mass for the galaxy, which in practice constrains each realization of the CMF.
As a consequence, all the previous discussion on how the choice of stopping criterion affect the \slug\ 
outputs in cluster simulations applies also to the way with which \slug\ approximates the 
galaxy mass by drawing from the CMF. Through the $f_{\rm c}$ parameter, the user has 
control on which level of this hierarchy contributes to the final simulation. In the limit 
of $f_{\rm c} = 0$, \slug\ removes the intermediate cluster container from the simulations: 
a galaxy mass is approximated by stars drawn from the IMF, without the constraints imposed
by the CMF. Conversely, in the limit of $f_{\rm c} = 1$, the input SFH constrains the shape of 
each realization of the CMF, which in turn shapes the mass spectrum of stars within the final 
outputs. As already noted in \citet{fumagalli11a}, this combined effect resembles in spirit 
the concept behind the IGIMF theory. However, the \slug\ implementation is fundamentally different
from the IGIMF, as our code does not require any \textit{a priori} modification of the functional form 
of the input IMF and CMF, and each realization of these PDFs is only a result of random sampling
of invariant PDFs. 

\red{To offer an example that better illustrates these concepts, we perform four ``galaxy'' simulations 
with 5000 realizations. This number ensures that $\sim 10$ simulations are present in each mass bin considered 
in our analysis, thus providing a well converged distribution}.
 Each simulation follows a single timestep of $2\times 10^6$ yr and assumes 
a \citet{chabrier05a} IMF, the default {\tt STOP\_NEAREST} condition, and an input SFR of 
0.001 $\rm M_\odot~yr^{-1}$. Three of the four simulations further assume a default CMF of the form
$dN/dM \propto M^{-2}$ between $20-10^7$ M$_\odot$, but 
three different choices of CMF ($f_{\rm c}=1.0$, 0.5, 
and 0.0). The fourth simulation still assumes $f_{\rm c}=1$ and a $dN/dM \propto M^{-2}$ CMF, 
but within the mass interval $20-100$ M$_\odot$ (hereafter the truncated CMF).
Results from these simulations are shown in \autoref{fig:clusterprop}.

By imposing an input SFR of 0.001 $\rm M_\odot~yr^{-1}$ for a time $2\times 10^6$ yr, we are 
in essence requesting that \slug\ populates galaxies with $2000$ M$_\odot$ worth of stars. 
However, similarly to the case of cluster simulations, 
\slug\ does not recover the input galaxy mass exactly, but it finds the best approximation 
based on the chosen stop criteria, the input CMF, and the choice of $f_{\rm c}$. 
This can be seen in the top panel of \autoref{fig:clusterprop}, where we show
the histograms of actual masses from the four simulations under consideration.
For $f_{\rm c} = 0$, \slug\ is approximating the target galaxy mass 
simply by drawing stars from the IMF. As the typical stellar mass is much less than the 
desired galaxy mass, the mass constraint is not particularly relevant in these simulations
and \slug\ reproduces the target galaxy mass quite well, as one can see from the narrow 
distribution centered around $2000$ M$_\odot$. When $f_{\rm c}>0$, however, \slug\ 
tries to approximate the target mass by means of much bigger building blocks, the clusters, 
thus increasing the spread of the actual mass distributions as seen for instance in the 
$f_{\rm c} = 0.5$ case. \red{For $f_{\rm c} > 0$, clusters as massive as $10^7$ M$_\odot$
are potentially accessible during draws and, as a consequence of the {\tt STOP\_NEAREST} condition, one can notice a 
wide mass distribution together with a non-negligible tail at very low (including zero) actual 
galaxy masses. Including a $10^7$ M$_\odot$ cluster to approximate a $2000$ M$_\odot$ galaxy would
in fact constitute a larger error than leaving the galaxy empty! The importance of
this effect obviously depends on how massive is the galaxy compared to the upper limit of the CMF.
In our example, the choice of an unphysically small  ``galaxy'' with $2000~\rm M_\odot$ 
is indeed meant to exacerbate the relevance of this mass constraint. 
By inspecting Figure  \ref{fig:clusterprop}, it is also evident that the resulting 
galaxy mass distributions are asymmetric, with a longer tail to lower masses. 
This is due to the fact that \slug\ favours small building blocks over massive ones
when drawing from a power-law CMF. This means that a draw of a cluster with 
mass comparable to the galaxy mass will likely occur after a few lower mass clusters
have been included in the stellar population. Therefore, massive clusters are more 
likely to be excluded that retained in a simulation. For this reason, the galaxy target mass
represents a limit inferior for the mean actual mass in each distribution.}

The middle panel of Figure  \ref{fig:clusterprop} shows the total amount of mass formed in 
clusters after one timestep. As expected, the fraction of mass in clusters versus field 
stars scales proportionally to $f_{\rm c}$, retaining the similar degree of 
dispersion noted for the total galaxy mass. Finally, by comparing the results of the 
$f_{\rm c} = 1$ simulations with the default CMF and the truncated CMF in all the three panels
of \autoref{fig:clusterprop}, one can appreciate the subtle difference that the choice of 
$f_{\rm c}$ and CMF have on the output. Even in the case of $f_{\rm c} = 1$, the truncated CMF 
still recovers the desired galaxy mass with high-precision. Obviously, this is an effect of the 
extreme choice made for the cluster mass interval, here between $20-100$ M$_\odot$. In this case, 
for the purpose of constrained sampling, the CMF becomes indistinguishable from the 
IMF, and the simulations of truncated CMF and $f_{\rm c} = 0$ both recover the target galaxy mass
with high accuracy. However, the simulations with truncated CMF still impose a constraint on 
the IMF, as shown in the bottom panel. In the case of truncated CMF, only clusters up to 
100 M$_\odot$ stars are formed, thus reducing the probability of drawing stars as massive as
120 M$_\odot$ from the IMF. 

This example highlights how the $f_{\rm c}$ parameter and the CMF need to be chosen 
with care based on the problem that one wishes to simulate, as they regulate in a
non-trivial way the scatter and the shape of the photometry distributions recovered by 
\slug. \red{In passing, we also note that \slug's ability to handle very 
general forms for the CMF and IMF makes our code suitable to explore a wide range of models
in which the galaxy SFR, CMF and IMF depend on each other \citep[e.g. as in the IGIMF;][]{weidner06a,weidner10a}.}

\begin{figure}
\includegraphics[width=\columnwidth]{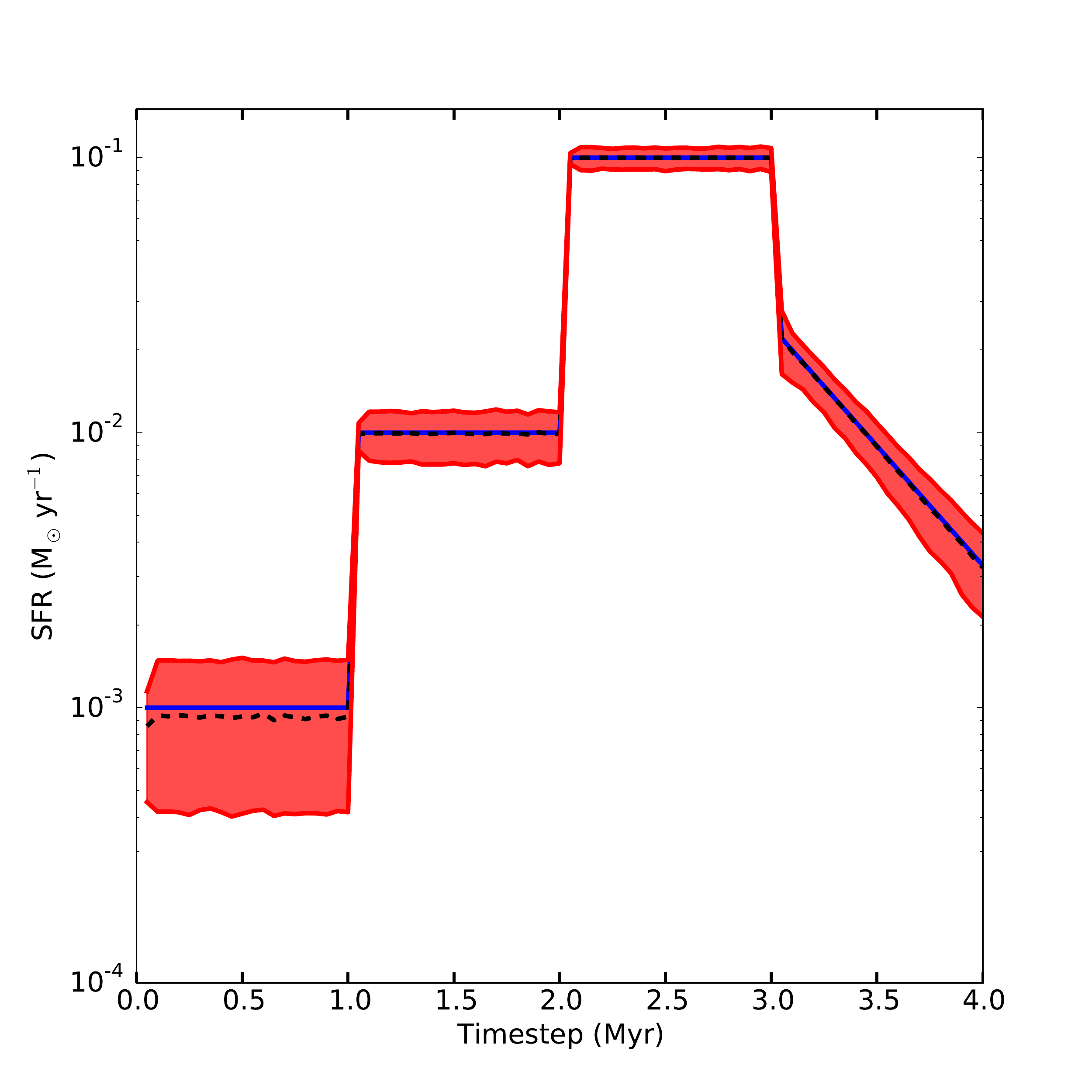}
\caption{\label{fig:sfhcomplex}
Realizations of the SFH \red{in 5000 \slug\ simulations} with $f_{\rm c}=1$. 
The input SFH is shown in blue, while the dashed black line and shaded regions 
show the median, and the first and third quartiles of the distribution.}
\end{figure}

\begin{figure*}
\includegraphics[scale=0.45]{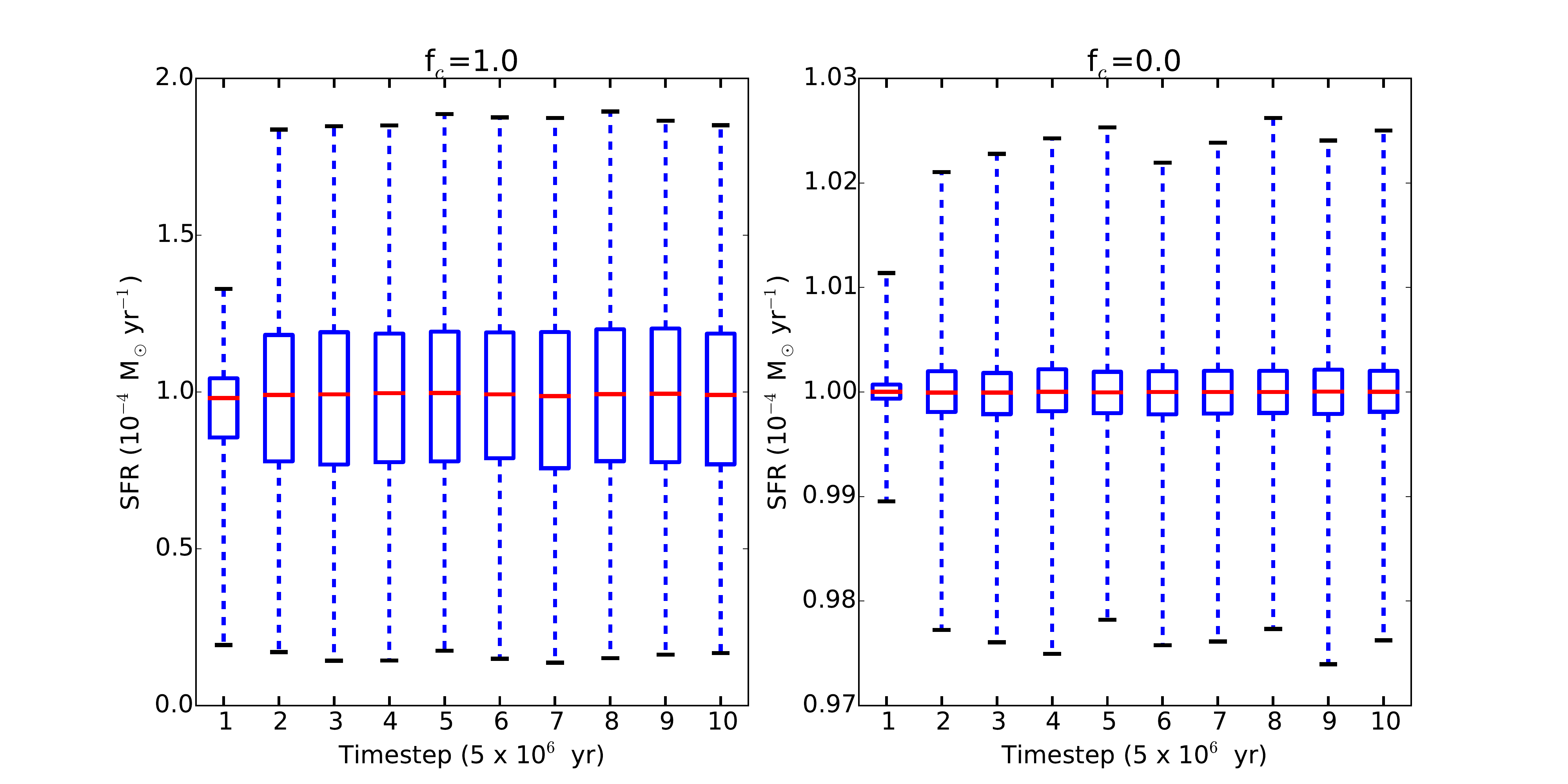}
\caption{\label{fig:sfhistory}
Whisker plots of the recovered SFH from  
two ``galaxy'' simlations \red{of 5000 trials each} and input constant SFR 
of $10^{-4}$ M$_\odot$ yr$^{-1}$. \red{The cluster fraction was set to 
$f_{\rm c} = 1.0$ (left panel) and $f_{\rm c} = 0.0$ (right panel)}.
At each timestep, the first and third quartiles are shown by the box plots, 
with the median marked by the red lines. The $5-$ and $95-$percentiles
are also shown by the whiskers.}
\end{figure*}

\subsection{Realizations of a Given Star Formation History}

\red{The study of SFHs in stochastic regimes is receiving 
much attention in the recent literature, both in the nearby and high-redshift 
universe
\citep[e.g.][]{kelson14a,dominguez14a,boquien14a}. As it can can handle arbitrary SFH in input, \slug\
is suitable for the analysis of stochastic effects on galaxy SFR.} 

In previous papers, and particularly in \citet{da-silva12a} and \citet{da-silva14b},
we have highlighted the conceptual difference between the input SFH and 
the outputs that are recovered from \slug\ simulations. 
The reason for such a difference should now be clear from the above discussion: 
\slug\ approximates an input SFH by means of discrete units, either in the 
form of clusters (for $f_{\rm c} = 1$), stars (for $f_{\rm c} = 0$), or a combination of both (for $0 < f_{\rm c} < 1$).
Thus, any smooth input function for the SFH (including a constant SFR)
is approximated by \slug\ as a series of bursts, that can described 
conceptually as the individual draws from the IMF or CMF. The effective SFH that \slug\ creates in
output is therefore an irregular function, which is the result of a 
superimposition of these multiple bursts. A critical ingredient is the 
typical time delays with which these bursts are combined, a quantity that is 
implicitly set by the SFH evaluated in each timestep and by the typical mass of 
the building blocks used to assemble the simulated galaxies. 

A simple example, which also highlights \slug's flexibility in handling 
arbitrary SFHs in input, is presented in \autoref{fig:sfhcomplex}. 
For this calculation, \red{we run 5000 \slug\ models} with default parameters and 
$f_{\rm c}=1$. The input SFH is defined by three segments of constant SFR 
across three time intervals of 1 Myr each, plus a fourth segment of
exponentially decaying SFR with timescale 0.5 Myr.  \autoref{fig:sfhcomplex} shows how
the desired SFH is recovered by \slug\ on average, but individual models show a substantial scatter about the 
mean, especially at low SFRs. An extensive discussion of this result is 
provided in section 3.2 of \citet{da-silva12a}.

Briefly, at the limit of many bursts and small time delays (i.e.
for high SFRs and/or when \slug\ populates galaxies 
mostly with stars for $f_{\rm c} \sim 0$), the output SFHs are reasonable 
approximations of the input SFH. Conversely, for small sets of bursts 
and for long time delays (i.e. for low SFRs and/or when \slug\ populates galaxies 
mostly with massive clusters for $f_{\rm c} \sim 1$), the output SFHs are only 
a coarse representation of the desired input SFH. 
This behaviour is further illustrated by \autoref{fig:sfhistory}, in which we 
show the statistics of the \red{SFHs of 5000 galaxy simulations}. These simulations
are performed assuming a constant SFR and two choices of fraction of stars 
formed in clusters, $f_{\rm c} =1$ and 0.
One can notice that, in both cases, a ``flickering'' SFH is recovered, but 
that a much greater scatter is evident for the $f_{\rm c} \sim 1$ case when
clusters are used to assemble the simulated galaxies.

 From this discussion, it clearly emerges that each \slug\ simulation will 
have an intrinsically bursty SFH, regardless to the user-set input, as already 
pointed out in \citet{fumagalli11a} and \citet{da-silva12a}.
\red{It is noteworthy that this fundamental characteristic associated to the 
discreteness with which star formation occurs in galaxies has also been 
highlighted by recent analytic and simulation work 
\citep[e.g.][]{kelson14a,dominguez14a,boquien14a,rodriguez-espinosa14a}. This effect, 
and the consequences it has on many aspects of galaxy studies
including the completeness of surveys or the use of SFR indicators,
is receiving great attention particularly in the context of studies at high redshift.
\Slug\ thus provides a valuable tool for further investigation into this problem,
particularly because our code naturally recovers a level of burstiness
imposed by random sampling, which does not need to be specified a-priori as  
in some of the previous studies.}

\begin{figure}
\includegraphics[width=\columnwidth]{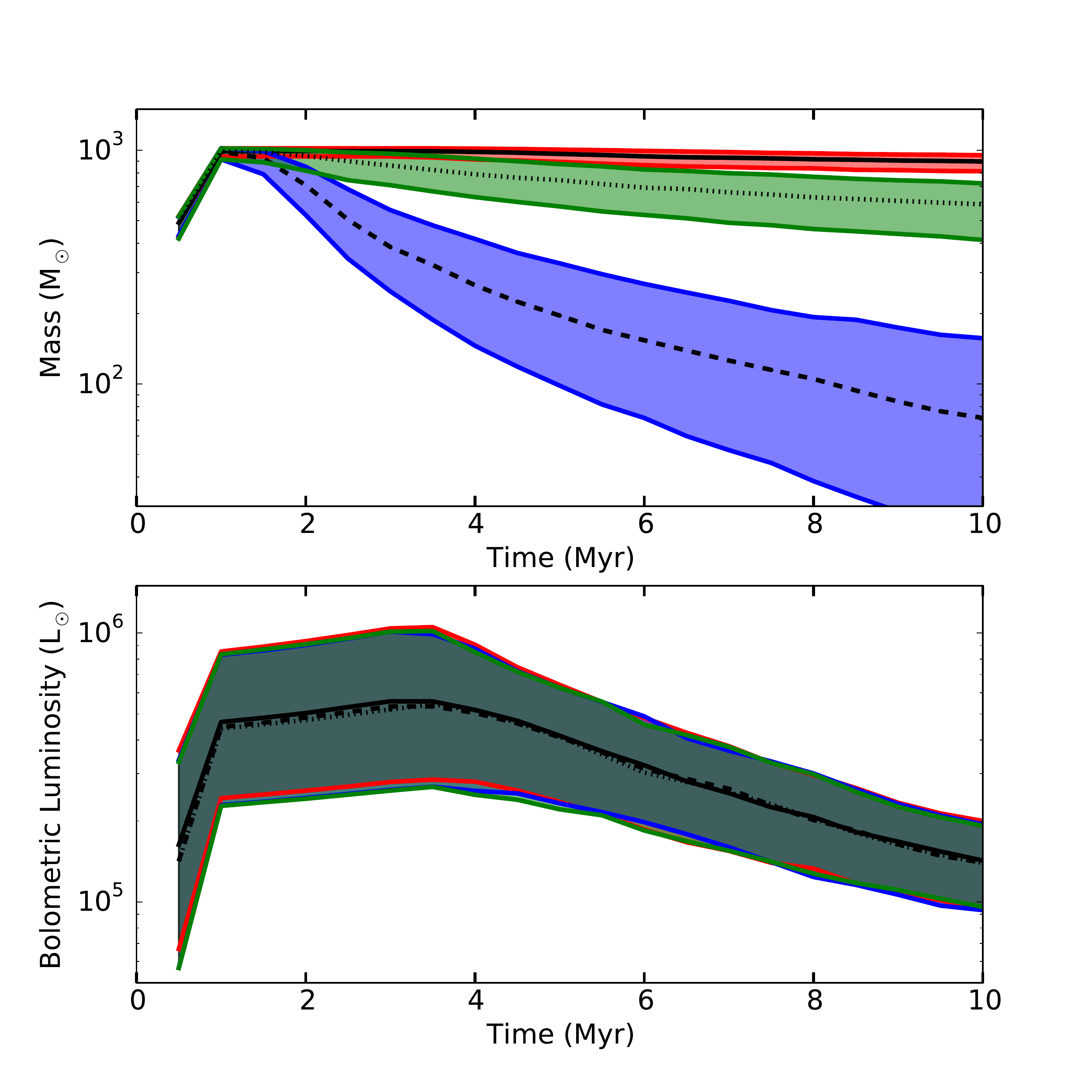}
\caption{\label{fig:clusterdis}
Total mass inside clusters (top) and the galaxy bolometric luminosity (bottom) 
as a function of time for three \slug\ simulations of \red{1000 trials} each.
In all cases, 1000 M$_\odot$ of stars are formed in a single burst within 1 Myr 
from the beginning of the calculation. Simulations without cluster disruption 
are shown in red, while simulations with cluster disruption enabled according to a
power-law CLF of index $-1.9$ and $-1$ are shown in blue and green.
The black and coloured thick lines show the median, first, and third quartiles of 
the distributions.}
\end{figure}

\subsection{Cluster Disruption}

When performing galaxy simulations, cluster disruption can be enabled in \slug.
In this new release of the code, \slug\ handles cluster disruption quite flexibly,
as the user can now specify the 
cluster lifetime function (CLF), which is a PDF 
from which the lifetime of each cluster is drawn.
This implementation generalises the original cluster disruption method
described in \citet{da-silva12a} to handle the wide range of lifetimes
observed in nearby galaxies (A.~Adamo et al., 2015, submitted).
We note however that the \slug\ default CLF 
still follows a power law of index $-1.9$ between 1 Myr and 1 Gyr 
as in \citet{fall09a}.

To demonstrate the cluster disruption mechanism, we run three simulations
of \red{1000 trials} each. These simulations follow the evolution of 
a burst of 1000 M$_\odot$ between $0-1$ Myr with a timestep of $5 \times 10^5$ yr  
up to a maximum time of 10 Myr. All stars are formed in clusters. 
The three simulations differ only for the choice of cluster disruption: one 
calculation does not implement any disruption law, while the other two 
assume a CLF in the form of a power law with indices $-1.9$ and $-1$
between $1-1000$ Myr. Results from these calculations 
are shown in \autoref{fig:clusterdis}.

The total mass in clusters, as expected, rises till a maximum of 1000 M$_\odot$
at 1 Myr, at which point it remains constant for the non-disruption case, while it 
declines according to the input power law in the other two simulations.
When examining the galaxy bolometric luminosity, one can see that the cluster disruption 
as no effect on the galaxy photometry. In this example, all stars are formed in clusters
and thus all the previous discussion on the mass constraint also applies here.
However, after formation, clusters and galaxies are passively evolved in \slug\ by 
computing the photometric properties as a function of time. When a cluster is disrupted,
\slug\ stops tagging it as a ``cluster'' object, but it still follows the contribution 
that these stars make to the integrated ``galaxy'' properties.
Clearly, \red{more complex examples in which star formation proceeds both in the 
field and in clusters following an input SFH while cluster disruption is 
enabled would exhibit photometric properties that are 
set by the passive evolution of previously-formed stars and by the 
zero-age main sequence properties of the newly formed stellar populations, each with its
own mass constraint}.

\begin{figure}
\includegraphics[width=\columnwidth]{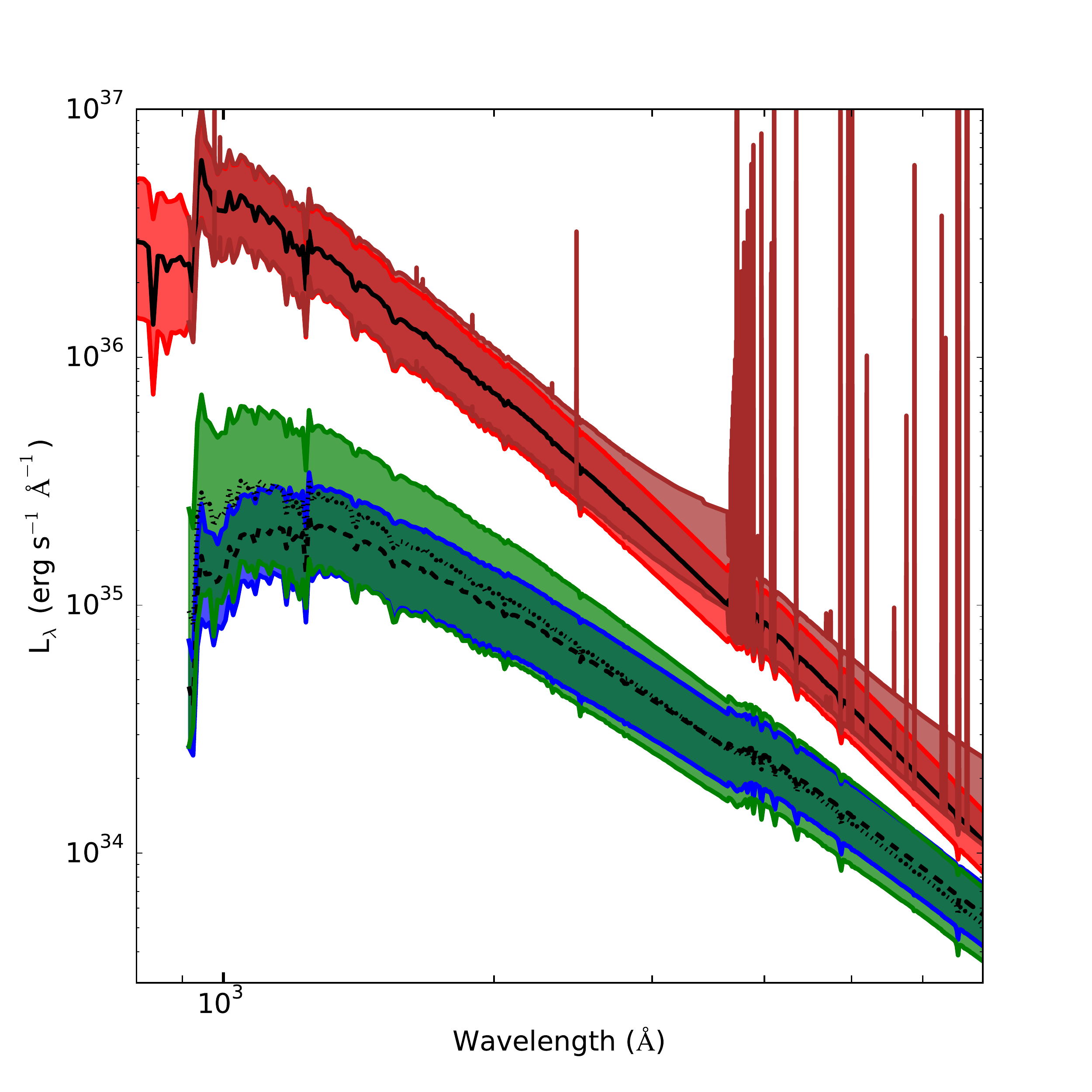}
\caption{\label{fig:specdust}
Spectra of four ``galaxy'' simulations with SFR of $0.001$ M$_\odot$ yr$^{-1}$
evaluated at time of $2\times 10^{6}$ yr. Each simulation differs for the adopted extinction
law or the inclusion of nebular emission. 
In red (solid line), the intrinsic stellar spectrum is shown, while models with deterministic 
and probabilistic extinction laws are shown respectively in blue
(dashed line) and green (dotted line). The solid lines show the first, second, and third
quartiles of \red{5000 realizations}. In brown, the range of luminosities (first and third quartiles) 
with the inclusion of nebular emission is shown.}
\end{figure}

\subsection{Dust Extinction and Nebular Emission}

In this section we offer an example of simulations which implement dust extinction and nebular emission
in post-processing, two new features offered starting from this release of the code 
(see \autoref{sssec:ppspectra}). \autoref{fig:specdust} shows the stellar spectra of three 
\slug\ simulations  (in green, blue, and red respectively) of a galaxy 
that is forming stars with a SFR of $0.001$ M$_\odot$ yr$^{-1}$. During these simulations,
\red{each of 5000 trials}, the cluster fraction is set to $f_{\rm c}=1$ and photometry 
is evaluated at $2\times 10^6$ yr. Three choices of extinction law are adopted: the first simulation 
has no extinction, while the second and third calculations implement the  \citet{calzetti00a} extinction law. 
In one case, a deterministic uniform screen with $A_{\rm V} = 1$ mag is applied to the emergent spectrum, 
while in the other case the \red{value of $A_{\mathrm{V}}$ is} drawn 
for each model from a lognormal distribution with mean 1.0 mag and dispersion 
of $\sim 0.3$ dex (the \slug\ default choice). 

As expected, the simulations with a deterministic uniform dust 
screen closely match the results of the non-dusty case, with a simple 
wavelength-dependent shift in logarithmic space. For the probabilistic 
dust model, on the other hand, the simulation results are qualitatively
similar to the non-dusty case,
but display a much greater scatter due to the varying degree of extinction that is 
applied in each trial. This probabilistic dust implementation allows one to more closely mimic the 
case of non-uniform dust extinction, in which different line of sights may be subject to 
a different degree of obscuration. 
One can also see how spectra with dust extinction are computed only for 
$\lambda >912$ \AA. This is not a physical effect, but it is a mere consequence of the 
wavelength range for which the extinction curves have been specified. 

\begin{figure*}
\includegraphics[scale=0.45]{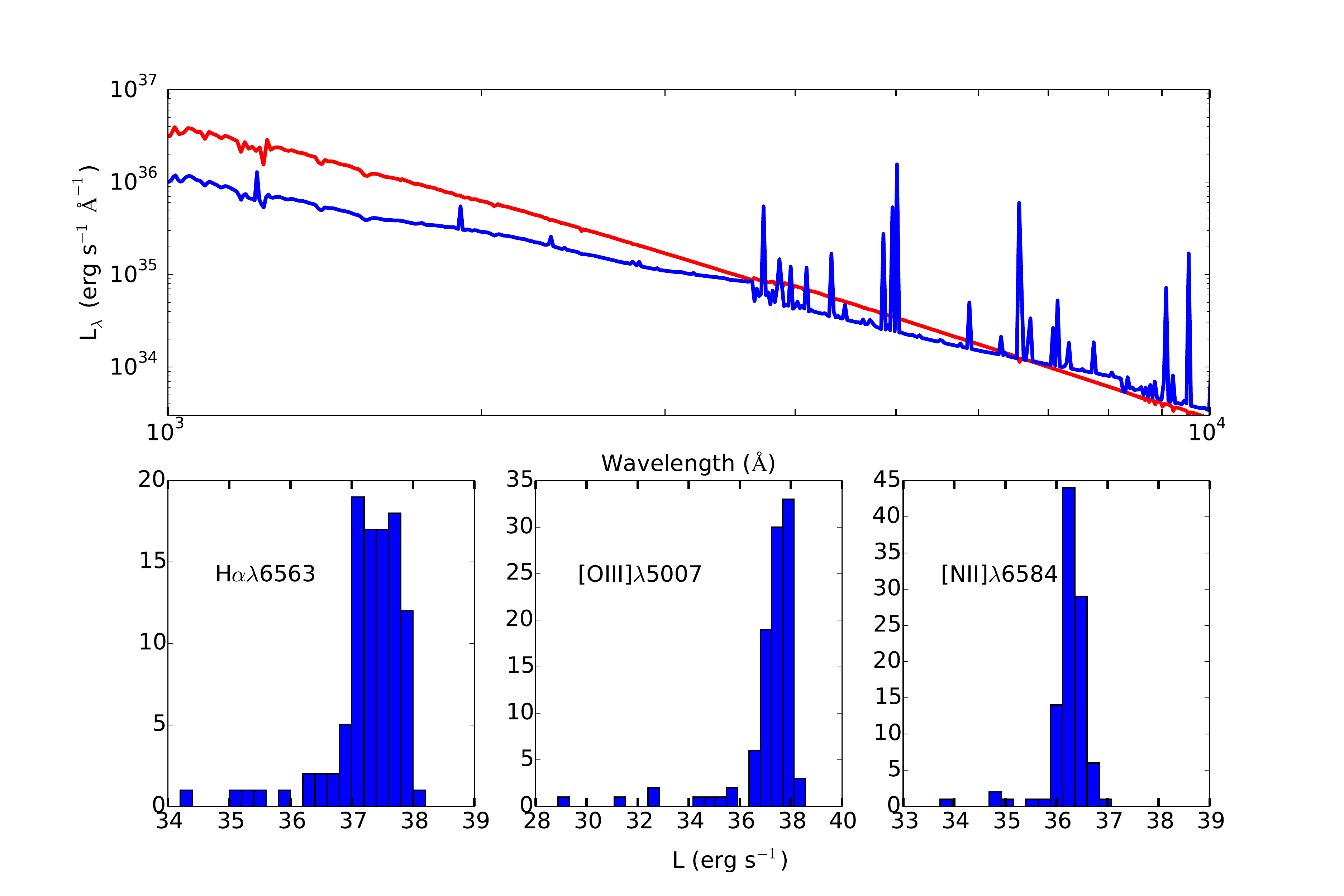}
\caption{\label{fig:cldy1}
Results of 100 \slug\ and {\tt cloudy} simulations for a galaxy with continuous 
SFR of 0.001 M$_\odot$ yr$^{-1}$ and $f_{\rm c}=1$. Top panel:
the median SED derived from the 100 \slug\ calculations is shown in red, 
while the median of the {\tt cloudy} SEDs computed in integrated mode is shown in blue.
Bottom panels: histograms of the line luminosity for three selected transitions computed 
by {\tt cloudy} in integrated mode for each input \slug\ SED.}
\end{figure*}

Finally, we also show in \autoref{fig:specdust} a fourth simulation  (brown colour), which is 
computed including nebular emission \red{with \slug's default parameters: $\phi=0.73$ and
$\log\,\mathcal{U} = -3$ (see Appendix \ref{app:nebuladust})}. In this case, the cutoff visible at the ionisation edge is physical, 
as \slug\ reprocesses the ionising radiation into lower frequency nebular emission according to the
prescriptions described in \autoref{sssec:ppspectra}. One can in fact notice,
besides the evident contribution from recombination lines, an increased luminosity at $\lambda \gtrsim 2500$ 
\AA\ that is a consequence of free-free, bound-free, and two-photon continuum emission.

\subsection{Coupling {\tt cloudy} to \slug\ Simulations}

In this section, we demonstrate the capability of the \cloudyslug\ package 
that inputs the results of \slug\ simulations into the 
{\tt cloudy} radiative transfer code. For this, we 
run 100 realizations of a ``galaxy'' simulation 
following a single timestep of $2\times 10^6$ yr. \red{The input parameters
for the simulation are identical to those in \autoref{ssec:clustfract}.}
The galaxy is forming 
stars at a rate of 0.001 M$_\odot$ yr$^{-1}$ with $f_{\rm c}=1$. We then 
pipe the \slug\ output into {\tt cloudy} to simulate H~\textsc{ii} regions in  
integrated mode, following the method discussed in \autoref{sec:cldy}. 
In these calculations, we assume the default parameters in \cloudyslug,
and in particular a density in the surroundings of the H~\textsc{ii} regions of $10^3~\rm cm^{-3}$.

Results are shown in \autoref{fig:cldy1}. In the top panel, 
the median of the 100 \slug\ SEDs is compared to the median of the 100 
SEDs returned by {\tt cloudy}, which adds the contribution of both the 
transmitted and the diffuse radiation. As expected, the processed 
{\tt cloudy} SEDs resemble the input \slug\ spectra. Photons at short 
wavelengths are absorbed inside the H~\textsc{ii} regions and are converted into 
UV, optical, and IR photons, which are then re-emitted within emission 
lines or in the continuum.\red{\footnote{\red{We do not show the region below 912
\AA~in \autoref{fig:cldy1} so that we can zoom-in on the features in the optical region.}}}
\red{The nebula-processed spectrum also shows the effects
of dust absorption within the H~\textsc{ii} region and its surrounding shell, which explains why
the nebular spectrum is suppressed below the intrinsic stellar one at short wavelengths.}

The bottom panel shows the full distribution of the line fluxes in three
selected transitions: H$\alpha$, [O~\textsc{iii}] $\lambda 5007$ and [N~\textsc{ii}] $\lambda 6584$.
These distributions highlight how the wavelength-dependent scatter in the input \slug\ SEDs
is inherited by the reprocessed emission lines, which exhibit a varying degree 
of stochasticity. \red{The long tail of the distribution at low H$\alpha$ luminosity is not
well-characterized by the number of simulations we have run, but this could be improved
simply by running a larger set of simulations. We are in the process of completing a much
more detailed study of stochasticity in line emission (T.~Rendahl et al., in preparation).} 

\subsection{Bayesian Inference of Star Formation Rates}

To demonstrate the capabilities of \sfrslug, we consider the simple example of using a measured ionizing photon flux to infer the true SFR. We use the library described above and in \citet{da-silva14b}, and consider ionizing fluxes which correspond to $\mathrm{SFR}_{Q(\mathrm{H}^0)} = 10^{-5}$, $10^{-3}$, and $10^{-1}$ $M_\odot$ yr$^{-1}$ using \red{an estimate that} neglects stochasticity and simply adopts the ionizing luminosity to SFR conversion appropriate for an infinitely-sampled IMF and SFH. We then use \sfrslug\ to compute the true posterior probability distribution on the SFR using these measurements; we do so on a grid of 128 points, using photometric errors of 0 and $0.5$ dex, and using two different prior probability distributions: one that is flat in $\log\mathrm{SFR}$ (i.e., $dp/d\log\mathrm{SFR} \sim \mathrm{constant}$), and one following the Schechter function distribution of SFRs reported by \citet{bothwell11a}, $dp/d\log\mathrm{SFR} \propto \mathrm{SFR}^{\alpha} \exp(-\mathrm{SFR}/\mathrm{SFR}_*)$, where $\alpha=-0.51$ and $\mathrm{SFR}_* = 9.2$ $M_\odot$ yr$^{-1}$.

\begin{figure}
\includegraphics[width=0.5\textwidth]{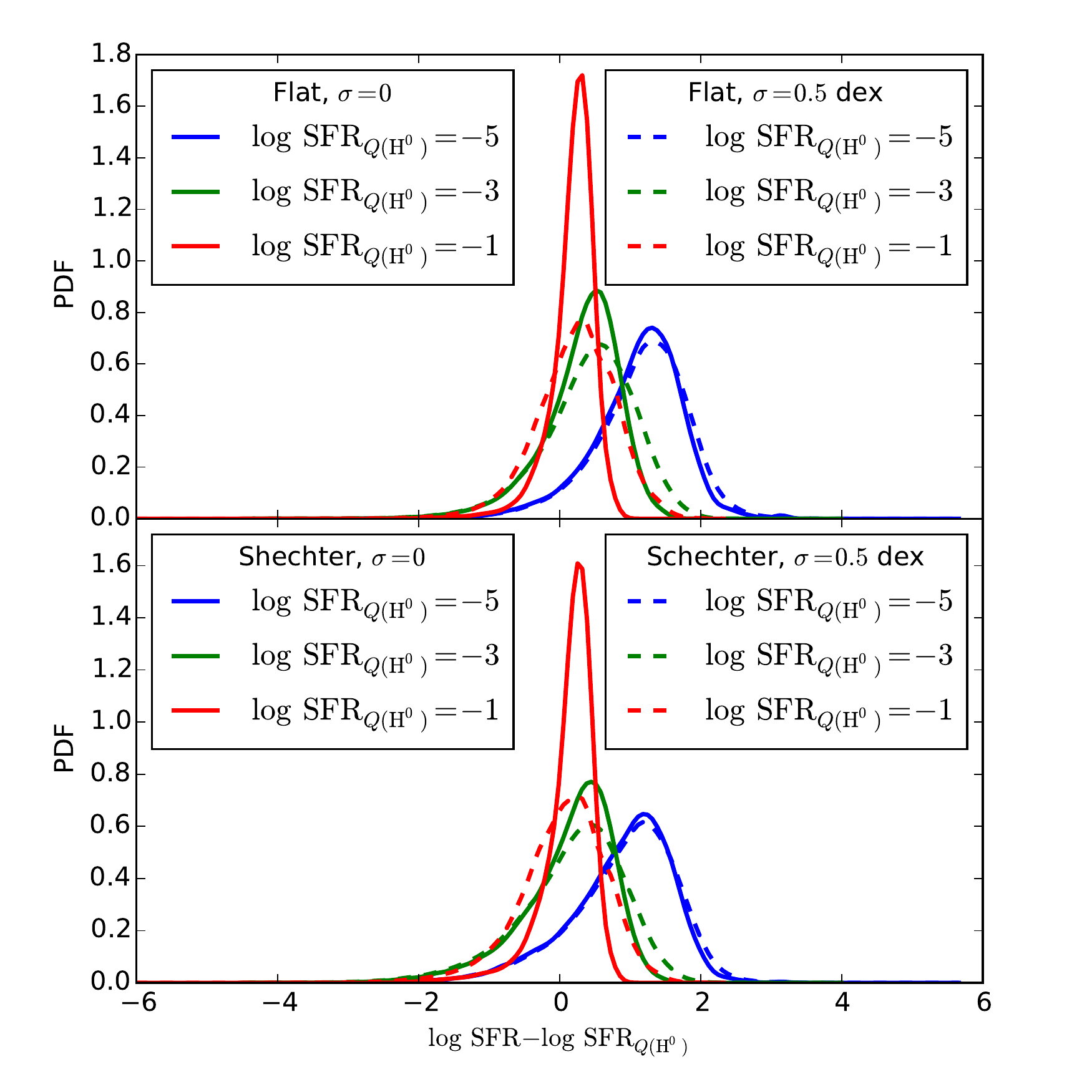}
\caption{
\label{fig:sfr_slug}
Posterior probability distributions for the logarithmic SFR based on measurements of the ionizing flux, computed with \sfrslug. The quantity plotted on the $x$ axis is the offset between the true $\log\mathrm{SFR}$ and the value that would be predicted by the ``point mass'' estimate where one ignores stochastic effects and simply uses the naive conversion factor between ionizing luminosity and SFR. Thus a value centered around zero indicates that the point mass estimate returns a reasonable prediction for the most likely SFR, while a value offset from zero indicates a systematic error in the point mass estimate. In the top panel, solid curves show the posterior PDF for a flat prior probability distribution and no photometric errors ($\sigma=0$), with the three colors corresponding to point mass estimates of $\log\mathrm{SFR}_{Q(\mathrm{H}^0)} = -5$, $-3$, and $-1$ based on the ionizing flux. The dashed lines show the same central values, but with assumed errors of $\sigma=0.5$ dex in the measured ionizing flux. In the bottom panel, we show the same quantities, but computed using a Schechter function prior distribution rather than a flat one (see main text for details).
}
\end{figure}

\autoref{fig:sfr_slug} shows the posterior PDFs we obtain, which we normalised to have unit integral. Consistent with the results reported by \citet{da-silva14b}, at SFRs of $\sim 10^{-1}$ $M_\odot$ yr$^{-1}$, the main effect of stochasticity is to introduce a few tenths of a dex uncertainty into the SFR determination, while leaving the peak of the probability distribution centered close to the value predicted by the point mass estimate. For SFRs of $10^{-3}$ or $10^{-5}$ $M_\odot$ yr$^{-1}$, the true posterior PDF is very broad, so that even with a $0.5$ dex uncertainty on the photometry, the uncertainty on the true SFR is dominated by stochastic effects. Moreover, the peak of the PDF differs from the value given by the point mass estimate by more than a dex, indicating a systematic bias. These conclusions are not new, but we note that the improved computational method described in Section~\ref{ssec:bp} results in a significant code speedup compared to the method presented in \citet{da-silva14b}. The time required for \sfrslug\ to compute the full posterior PDF for each combination of $\mathrm{SFR}_{Q(\mathrm{H}^0)}$, photometric error, and prior probability distribution is $\sim 0.2$ seconds on a single core of a laptop (excluding the startup time to read the library). Thus this method can easily be used to generate posterior probability distributions for large numbers of measurement in times short enough for interactive use.

\subsection{Bayesian Inference of Star Cluster Properties}

To demonstrate the capabilities of \clusterslug, we re-analyse the catalog of star clusters in the central regions of M83 described by \citet{chandar10b}. These authors observed M83 with \textit{Wide Field Camera 3} aboard the \textit{Hubble Space Telescope}, and obtained measurements for $\sim 650$ star clusters in the filters F336W, F438W, F555W, F814W, and F657N (H$\alpha$). They used these measurements to assign each cluster a mass and age by comparing the observed photometry to simple stellar population models using \citet{bruzual03a} models for a twice-solar metallicity population, coupled with a Milky Way extinction law; see \citet{chandar10b} for a full description of their method. \red{This catalog has also been re-analyzed by \citet{fouesneau12a} using their stochastic version of \texttt{pegagse}. Their method, which is based on $\chi^2$ minimization over a large library of simulated clusters, is somewhat different than our kernel density-based one, but should share a number of similarities -- see \citet{fouesneau10a} for more details on their method. We can therefore compare our results to theirs as well.}

We downloaded \citeauthor{chandar10b}'s ``automatic'' catalog from MAST\footnote{\url{https://archive.stsci.edu/}} and used \clusterslug\ to compute posterior probability distributions for the mass and age of every cluster for which photometric values were available in all five filters. We used the photometric errors included in the catalog in this analysis, coupled to the default choice of bandwidth in \clusterslug, and a prior probability distribution that is flat in the logarithm of the age and $A_V$, while varying with mass as $p(\log M)\propto 1/M$. \red{We used our library computed for the Padova tracks and a Milky Way extinction curve, including nebular emission.} The total time required for \clusterslug\ to compute the marginal probability distributions of mass and age on grids of 128 logarithmically-spaced points each was \red{$\sim 4$} seconds per marginal PDF (\red{$\sim 6000$} seconds for 2 PDFs each on the entire catalog of 656 clusters), using a single CPU. The computation can be parallelized trivially simply by running multiple instances of \clusterslug~on different parts of the input catalog.

\begin{figure}
\includegraphics[width=0.5\textwidth]{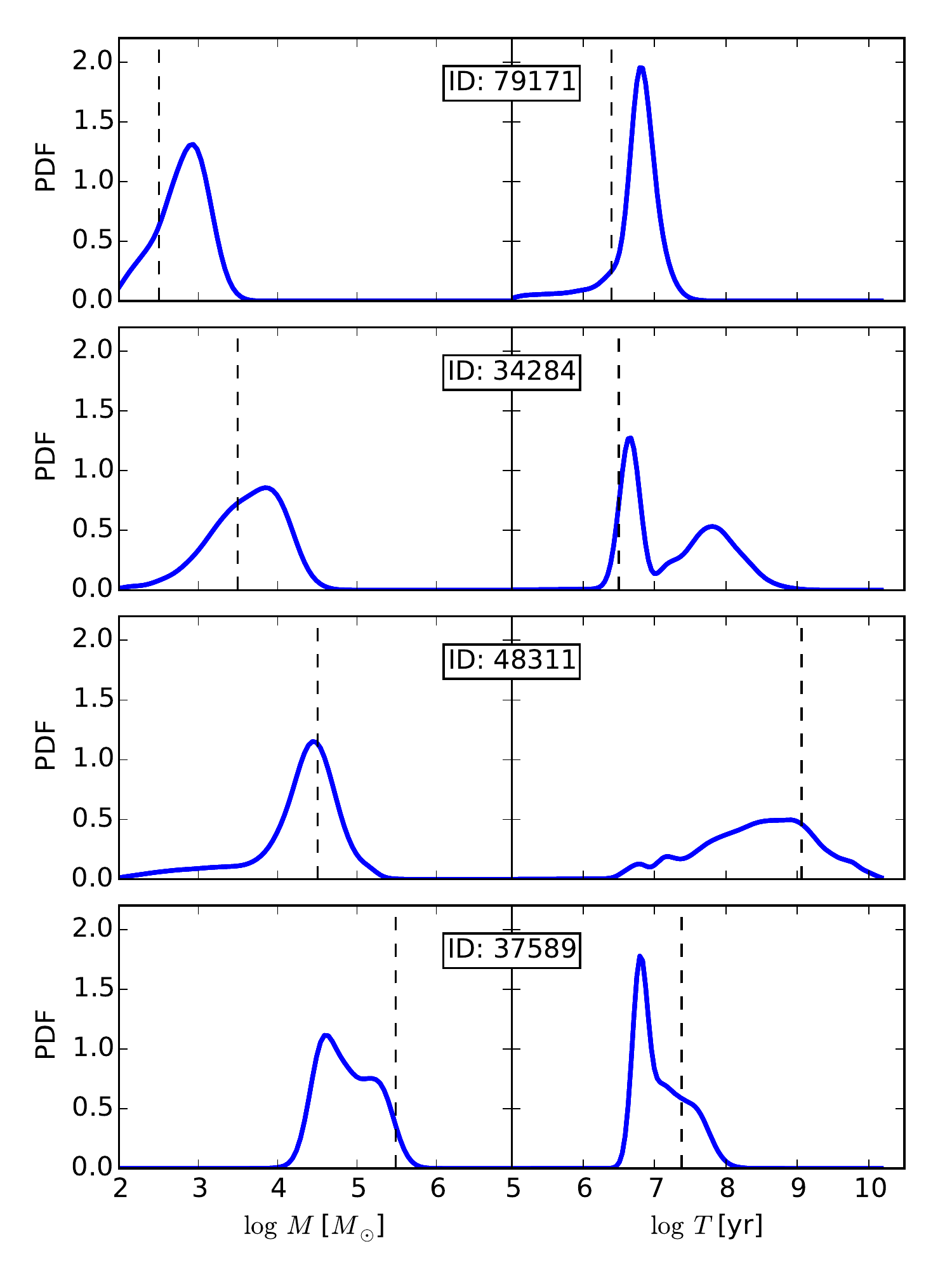}
\caption{\label{fig:cluster_slug_example_1}
Posterior probability distributions for cluster mass (left) and age (right) for four sample clusters taken from the catalog of \citet{chandar10b}, as computed by \clusterslug. In each panel, the heavy blue line is the \clusterslug~result, and the thin vertical dashed line is the best fit obtained by \citeauthor{chandar10b}. The ID number printed in each pair of panels is the  ID number assigned in \citeauthor{chandar10b}'s catalog.}
\end{figure}

\begin{figure*}
\includegraphics[width=\textwidth]{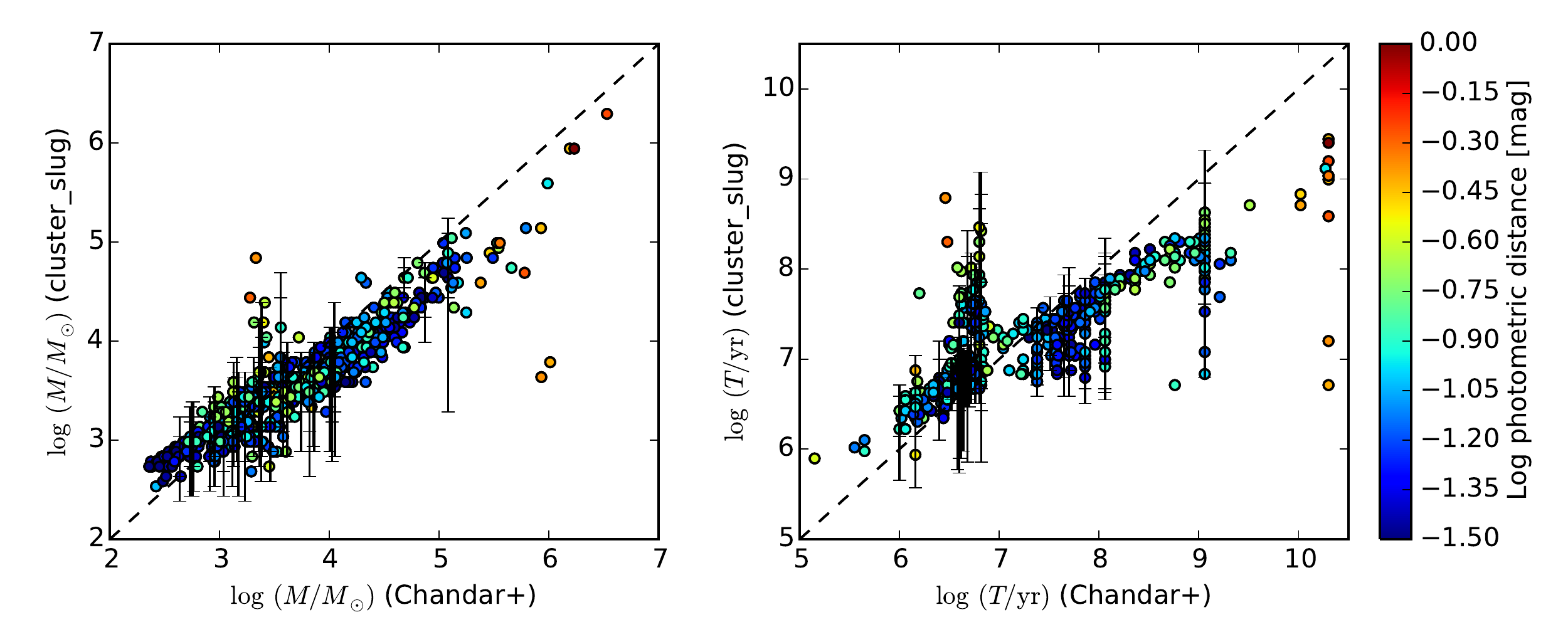}
\caption{
\label{fig:cluster_slug_example_2}
Comparison of cluster masses (left) and ages (right) for clusters in the \citet{chandar10b} catalog computed using two different methods. For each cluster, the value shown on the $x$ axis is the best-fit value reported by \citeauthor{chandar10b} using their non-stochastic model grids. The values shown on the $y$ axis are the median masses and ages computed by \clusterslug; for every 20th cluster we also show error bars, which indicate the \red{10th - 90th} percentile range computed from the \clusterslug~posterior PDF. The dashed black lines indicate the $1-1$ line, i.e., perfect agreement. Points are colored by the photometric distance (see main text for definition) between the observed photometry for that cluster and the 5th closest match in the \clusterslug~simulation library. 
}
\end{figure*}

In \autoref{fig:cluster_slug_example_1} we show example posterior probabilities distributions for cluster mass and age as returned by \clusterslug, and in \autoref{fig:cluster_slug_example_2} we show the median and interquartile ranges for cluster mass and age computed from \clusterslug~compared to those determined by \citet{chandar10b}\footnote{The mass and age estimates plotted are from the most up-to-date catalog maintained by R.~Chandar (2015, priv.~comm.).}. The points in \autoref{fig:cluster_slug_example_2} are color-coded by the ``photometric distance" between the observed photometric values and the 5th closest matching model in the \clusterslug~library, where the photometric distance is defined as
\begin{equation}
d = \sqrt{\frac{1}{N_{\mathrm{filter}}}\sum_{i=1}^{N_{\mathrm{filter}}} (M_{i,\mathrm{obs}} - M_{i,\mathrm{lib}})^2}
\end{equation}
where $M_{i,\mathrm{obs}}$ and $M_{i,\mathrm{lib}}$ are the observed magnitude and the magnitude of the \clusterslug~simulation in filter $i$, and the sum is over all 5 filters used in the analysis.

We can draw a few conclusions from these plots. Examining \autoref{fig:cluster_slug_example_1}, we see that \clusterslug~in some cases identifies a single most likely mass or age with a fairly sharp peak, but in other cases identifies multiple distinct possible fits, so that the posterior PDF is bimodal. In these cases the best fit identified by \citeauthor{chandar10b}~usually matches one of the peaks found by \clusterslug. The ability to recover bimodal posterior PDFs represents a distinct advantage of \clusterslug's method, since it directly indicates cases where there is a degeneracy in possible models.

From \autoref{fig:cluster_slug_example_2}, we see that, with the exception of a few catastrophic outliers, the median masses returned by \clusterslug~are comparable to those found by \citeauthor{chandar10b} The ages \red{show general agreement, but less so than the masses. For the ages, disagreements come in two forms. First, there is a systematic difference that \clusterslug~tends to assign younger ages for any cluster for which \citeauthor{chandar10b} have assigned an age $>10^8$ yr. For many of these clusters the $1-1$ line is still within the 10th - 90th percentile range, but the 50th percentile is well below the $1-1$ line. The second type of disagreement is more catastrophic. We find that there are are a fair number of clusters for which \citeauthor{chandar10b} assign ages of $\sim 5$ Myr, while \clusterslug~produces ages $1-2$ dex larger. Conversely, for the oldest clusters in \citeauthor{chandar10b}'s catalog, \clusterslug~tends to assign significantly younger ages.}

\red{The differences in ages probably have a number of causes. The fact that our 50th percentile ages tend to be lower than those derived by \citet{chandar10b} even when the disagreement is not catastrophic is likely a result of the broadness of the posterior PDFs, and the fact that we are exploring the full posterior PDF rather than selecting a single best fit. For example, in the middle two panels of \autoref{fig:cluster_slug_example_1}, the peak of the posterior PDF is indeed close to the value assigned by \citeauthor{chandar10b} However, the PDF is also very broad, so that the 50th percentile can be displaced very significantly from the peak. We note that this also likely explains why our \autoref{fig:cluster_slug_example_2} appears to indicate greater disagreement between stochastic and non-stochastic models than do \citet{fouesneau12a}'s Figures 18 and 20, which ostensibly make the same comparison. \citeauthor{fouesneau12a} identify the peak in the posterior PDF, but do not explore its full shape. When the PDF is broad or double-peaked, this can be misleading.}

\red{The catastrophic outliers are different in nature, and can be broken into two categories. The disagreement at the very oldest ages assigned by \citet{chandar10b} is easy to understand: the very oldest clusters in M83 are globular clusters with substantially sub-Solar metallicities, while our library is for Solar metallicity. Thus our library simply does not contain good matches to the colors of these clusters, as is apparent from their large photometric distances (see more below). The population of clusters for which \citet{chandar10b} assign $\sim 5$ Myr ages, while the stochastic models assign much older ages, is more interesting. \citet{fouesneau12a} found a similar effect in their analysis, and their explanation of the phenomenon holds for our models as well. These clusters have relatively red colors and lack strong H$\alpha$ emission, properties that can be matched almost equally well either by a model at an age of $\sim 5$ Myr (old enough for ionization to have disappeared) with relatively low mass and high extinction, or by a model at a substantially older age with lower extinction and higher total mass. This is a true degeneracy, and the stochastic and non-stochastic models tend to break the degeneracy in different directions. The ages assigned to these clusters also tend to be quite dependent on the choice of priors (Krumholz et al., 2015, in preparation).}

We can also see from \autoref{fig:cluster_slug_example_2} that, for the most part and without any fine-tuning, the default \clusterslug~library does a very good job of matching the observations. As indicated by the color-coding, for most of the observed clusters, the \clusterslug~library contains at least 5 simulations that match the observations to a photometric distance of a few tenths of a magnitude. Quantitatively, 90\% of the observed clusters have a match in the library that is within \red{0.15} mag, and 95\% have a match within \red{0.20} mag; for 5th nearest neighbors, these figures rise to 0.18 and \red{0.22} mag. There are, however, some exceptions, and these are clusters for which the \clusterslug~fit differs most dramatically from the \citeauthor{chandar10b} one. \red{The worst agreement is found for the clusters for which \citet{chandar10b} assigns the oldest ages, and, as noted above, this is almost certainly a metallicity effect. However,} even eliminating these cases there are $\sim 10$ clusters for which the closest match in the \clusterslug~library is at a photometric distance of $0.3-0.6$ mag. It is possible that these clusters have extinctions $A_V > 3$ and thus outside the range covered by the library, that these are clusters where our failure to make the appropriate aperture corrections makes a large difference, or that the disagreement has some other cause. These few exceptions notwithstanding, this experiment makes it clear that, for the vast majority of real star cluster observations, \clusterslug~can return a full posterior PDF that properly accounts for stochasticity and other sources of error such as age-mass degeneracies, and can do so at an extremely modest computational cost.

\section{Summary and Prospects}
\label{sec:summary}

As telescopes gains in power, observations are able to push to ever smaller spatial and mass scales. Such small scales are often the most interesting, since they allow us to observe fine details of the process by which gas in galaxies is transformed into stars. However, taking advantage of these observational gains will require the development of a new generation of analysis tools that dispense with the simplifying assumptions that were acceptable for processing lower-resolution data. The goal of this paper is to provide such a next-generation analysis tool that will allow us to extend stellar population synthesis methods beyond the regime where stellar initial mass functions (IMFs) and star formation histories (SFHs) are well-sampled. The \slug~code we describe here makes it possible to perform full SPS calculations with nearly-arbitrary IMFs and SFHs in the realm where neither distribution is well-sampled; the accompanying suite of software tools makes it possible to use libraries of \slug~simulations to solve the inverse problem of converting observed photometry to physical properties of stellar populations outside the well-sampled regime. In addition to providing a general software framework for this task, \bp, we provide software to solve the particular problems of inferring the posterior probability distribution for galaxy star formation rates (\sfrslug) and star cluster masses, ages, and extinctions (\clusterslug) from unresolved photometry.

In upcoming work, we will use \slug~and its capability to couple to \texttt{cloudy} \citep{ferland13a} to evaluate the effects of stochasticity on emission line diagnostics such as the BPT diagram \citep{baldwin81a}, and to analyze the properties of star clusters in the new Legacy Extragalactic UV Survey \citep{calzetti15a}. However, we emphasize that \slug, its companion tools, and the pre-computed libraries of simulations we have created are open source software, and are available for community use on these and other problems.

\section*{Acknowledgements}

The authors thank Michelle Myers for assistance, and Aida Wofford, \red{Angela Adamo, and Janice Lee} for comments on the manuscript. MRK acknowledges support from Hubble Space Telescope programs HST-AR-13256 and HST-GO-13364, provided by NASA through a grant from the Space Telescope Science Institute, which is operated by the Association of Universities for Research in Astronomy, Inc., under NASA contract NAS 5-26555. MF acknowledges support by the Science and Technology Facilities Council, grant number  ST/L00075X/1. Some of the data presented in this paper were obtained from the Mikulski Archive for Space Telescopes (MAST). STScI is operated by the Association of Universities for Research in Astronomy, Inc., under NASA contract NAS5-26555. Support for MAST for non-HST data is provided by the NASA Office of Space Science via grant NNX13AC07G and by other grants and contracts. This work used the \red{UCSC supercomputer Hyades,
which is supported by the NSF (award number AST-1229745), and the} DiRAC Data Centric system at 
Durham University, operated by the Institute for Computational Cosmology on behalf of the STFC 
DiRAC HPC Facility (www.dirac.ac.uk). This equipment was funded by BIS National E-infrastructure 
capital grant ST/K00042X/1, STFC capital grant ST/H008519/1, and STFC DiRAC Operations grant 
ST/K003267/1 and Durham University. DiRAC is part of the National E-Infrastructure. 

\bibliographystyle{mn2e}
\bibliography{refs}

\begin{appendix}

\section{\red{Implementation Details and Capabilities}}
\label{app:implementation}

\red{Here we describe some details of \slug's current implementation. These capabilities may be expanded in future versions, but we include this description here both to demonstrate the code's flexibility, and to discuss some subtleties that may be of use to those interested in implementing similar codes.}

\subsection{\red{Probability Distribution Functions}}
\label{app:pdfs}

\red{\Slug~uses a large number of probability distribution functions (PDFs). In particular, the IMF, the cluster mass function (CMF), and the the star formation history (SFH) are all required to be PDFs, and the extinction, output time, and (for simulations of simple stellar populations) can optionally be described by PDFs as well. In \slug, PDFs}, can be specified as a sum of an arbitrary number of piecewise continuous segments,
\begin{equation}
\label{eq:pdfforms}
\frac{dp}{dx} = n_1 f_1(x; x_{1,a}, x_{1,b}) + n_2 f_2(x; x_{2,a}, x_{2,b}) + \cdots,
\end{equation}
where the normalizations $n_i$ for each segment are free parameters, as are the parameters $x_{i,a}$ and $x_{i,b}$ that denote the lower and upper limits for each segment (i.e., the function $f_i(x)$ is non-zero only in the range $x \in [x_{i,a}, x_{i,b}]$). The functions $f_i(x)$ can take any of the functional forms listed in \autoref{tab:pdfforms}, and the software is modular so that additional functional forms can be added very easily. In the most common cases, the segment limits and normalizations will be chosen so that the segments are contiguous with one another and the overall PDF continuous, i.e., $x_{i,a} = x_{i,b}$ and $n_i f_i(x_{i,b}) = n_{i+1} f_{i+1}(x_{i+1,a})$. However, this is not a required condition, and the limits, normalizations, and number of segments can be varied arbitrarily. \red{In particular, segments are allowed to overlap or to be discontinuous (as they must in the case of $\delta$ function segments); thus for example one could treat the star formation history of a galaxy as a sum of two components, one describing the bulge and one describing the disc.} \Slug\ ships with the following IMFs pre-defined for user convenience: \citet{salpeter55a}, \citet{kroupa02c}, \citet{chabrier03a}, and \citet{chabrier05a}.

\begin{table}
 \centering
 \begin{minipage}{80mm}
  \caption{Functional forms for PDF segments in \slug. \label{tab:pdfforms}}
\begin{tabular}{@{}lcc@{}}
\hline
Name & Functional form, $f(x)$ & Parameters\footnote{In addition to the segment-specific parameters listed, all segments also allow the upper and lower cutoffs $x_a$ and $x_b$ as free parameters.} \\
\hline
\texttt{delta} & $\delta(x - x_0)$ & $x_0$\footnote{For \texttt{delta} segments, we require that $x_a = x_b = x_0$.} \\
\texttt{exponential} & $e^{-x/x_*}$ & $x_*$ \\
\texttt{lognormal} & $x^{-1} \exp[-(\ln x/x_0)^2/(2 s^2)]$ & $x_0$, $s$ \\
\texttt{normal} & $\exp[-(x-x_0)^2/(2 s^2)]$ & $x_0$, $s$ \\
\texttt{powerlaw} & $x^p$ & $p$ \\
\texttt{schechter} & $x^p e^{-x/x_*}$ & $p$, $x_*$\\
\hline
\end{tabular}
\end{minipage}
\end{table}

\red{When drawing a finite total mass, in addition to the mass distribution, one must also specify a sampling method to handle the fact that one will not be able to hit the target mass perfectly when drawing discrete objects.} \Slug\ allows users to choose a wide range of methods for this purpose, which we describe briefly here following the naming convention used in the code. \red{The majority of these methods are described in \citet{haas10a}.}

\begin{itemize}
\item \texttt{STOP\_NEAREST}: draw from the IMF until the total mass of the population exceeds $M_{\rm target}$. Either keep or exclude the final star drawn depending on which choice brings the total mass closer to the target value. 
Unless a different scheme is deemed necessary, this is the preferred and default choice of 
\slug, as this sampling technique ensures that the stochastic solution converges towards 
the deterministic one at the limit of sufficiently large $M_{\rm target}$.
\item \texttt{STOP\_BEFORE}: same as \texttt{STOP\_NEAREST}, but the final star drawn is always excluded.
\item \texttt{STOP\_AFTER}: same as \texttt{STOP\_NEAREST}, but the final star drawn is always kept.
\item \texttt{STOP\_50}: same as \texttt{STOP\_NEAREST}, but keep or exclude the final star with 50\% probability regardless of which choice gets closer to the target.
\item \texttt{NUMBER}: draw exactly $N = M_{\rm target}/\langle M\rangle$ objects, where $\langle M\rangle$ is the expectation value for the mass of an object produced by a single draw\red{, and the value of $N$ is rounded to the nearest integer. Note that this method can be used to handle the case of characterizing a population as containing a particular number of objects as opposed to a particular total mass, simply by choosing $M_{\rm target} = N \langle M \rangle$.}
\item \texttt{POISSON}: draw exactly $N$ objects, where the value of $N$ is chosen from a Poisson distribution with expectation value $\langle N \rangle = M_{\rm target}/\langle M\rangle$
\item \texttt{SORTED\_SAMPLING}\footnote{This method replaces the \texttt{IGIMF} method implemented in the earlier version of \slug\ \citep{da-silva12a}, which was based on the \citet{weidner10a} version of the integrated galactic IMF (IGIMF) model. In \citet{weidner10a}'s formulation, the upper cutoff of the IMF in a star cluster depends explicitly (rather than simply due to size-of-sample effects) on the total mass of the cluster. This model has been fairly comprehensively excluded by observations since the original \slug\ code was developed \citep{fumagalli11a, andrews13a, andrews14a}, and \citet{weidner14a} advocated dropping that formulation of the IGIMF in favor of the earlier \citet{weidner06a} one. See \citet{krumholz14c} for a recent review discussing the issue.}: this method was introduced by \citet{weidner06a}. In it, one first draws exactly $N= M_{\rm target}/\langle M\rangle$ objects as in the \texttt{NUMBER} method. If the resulting total mass $M_{\rm pop}$ is less than $M_{\rm target}$, the procedure is repeated recursively using a target mass $M_{\rm target} - M_{\rm pop}$ until $M_{\rm pop} > M_{\rm target}$. Finally, one sorts the resulting list of objects from least to most massive, and then keeps or removes the final, most massive using a \texttt{STOP\_NEAREST} policy. 
\end{itemize}

\red{Finally, we note two limitations in our treatment of the IMF. First, while \slug~allows a great deal of flexibility in its treatment of PDFs, it requires that the various PDFs that appear in the problem (IMF, CMF, etc.) be separable, in the sense that the one cannot depend on the other. Thus for example it is not presently possible to have an IMF that varies systematically over the star formation history of a simulation. Second, while the IMF can extend to whatever mass is desired, the ability of the code to calculate the light output depends on the availability of stellar evolution tracks extending up to the requisite mass. The set of tracks available in the current version of \slug~(Appendix \ref{app:tracks}) does not extend above $120$ $M_\odot$.}

\subsection{\red{Tracks and Atmospheres}}
\label{app:tracks}

The evolutionary tracks \red{used by \slug} consist of a rectangular grid of models for stars' present day-mass, luminosity,  effective temperature, and surface abundances at a series of times for a range of initial masses; the times at which the data are stored are chosen to represent equivalent evolutionary stages for stars of different starting masses, and thus the times are not identical from one track to the next. \Slug\ uses the same options for evolutionary tracks as \texttt{starburst99} \citep{leitherer99a, vazquez05a, leitherer10a, leitherer14a}\red{, and in general its treatment of tracks and atmospheres clones that of \texttt{starburst99} except for minor differences in the numerical schemes used for interpolation and numerical integration (see below)}. In particular, \slug\ implements the latest Geneva models for non-rotating and rotating stars \citep{ekstrom12a, georgy13a}, as well as earlier models from the Geneva and Padova groups \citep{schaller92a, meynet94a, girardi00a}\red{; the latter can also include a treatment of thermally pulsing AGB stars from \citep{vassiliadis93a}. The Geneva models are optimized for young stellar populations and likely provide the best available implementation for them, but they have a minimum mass of $0.8$ $M_\odot$, and do not include TP-AGB stars, so they become increasingly inaccurate at ages above $\sim 10^8$ yr. The Padova tracks should be valid up to the $\sim 15$ Gyr age of the Universe, but are less accurate than the Geneva ones at the earliest ages. See \citet{vazquez05a} for more discussion.} Models are available at a range of metallicities\red{; at present the latest Geneva tracks are available at $Z=0.014$ and $Z=0.002$, the older Geneva tracks are available at $Z=0.001$, 0.004, 0.008, 0.020, and 0.040, while the Padova tracks are available at $Z=0.0004$, 0.004, 0.008, 0.020, and 0.050.}

\Slug\ interpolates on the evolutionary tracks using a somewhat higher-order version of the isochrone synthesis technique \citep{charlot91a} adopted in most other SPS codes. The interpolation procedure is as follows: first, \slug\ performs \citet{akima70a} interpolation in both the mass and time directions for all variables stored on the tracks\red{; interpolations are done in log-log space. Akima interpolation is a cubic spline method with an added limiter that prevents ringing in the presence of outliers; it requires five points in 1D. For tracks where fewer than five points are available, \slug~reverts to linear interpolation.} To generate an isochrone, the code interpolates every mass and time track to the desired time, and then uses the resulting points to generate a new Akima interpolation at the desired time.

\red{Note that the choice of interpolation method does not affect most quantities predicted by SPS methods, but it does affect those that are particularly sensitive to discontinuities in the stellar evolution sequence. For example, the production rate of He$^+$-ionizing photons is particularly sensitive to the presence of WR stars, and thus to the interpolation method used to determine the boundary in mass and time of the WR region. We have conducted tests comparing \slug~run with stochasticity turned off to \texttt{starburst99}, which uses quadratic interpolation, and find that the spectra produced are identical to a few percent at all wavelengths, with the exception of wavelengths corresponding to photon energies above a few Rydberg at ages of $\sim 4$ Myr. Those differences trace back to differences in determining which stars are WR stars, with \slug's Akima spline giving a slightly different result that \texttt{starburst99}'s quadratic one. For a more extensive discussion of interpolation uncertainties in SPS models, see \citet{cervino05a}.}

Stellar atmospheres are also treated in the same manner as is in \texttt{starburst99} \red{\citep{smith02a}}. By default, stars classified as Wolf-Rayet stars based on their masses, effective temperatures, and surface abundances are handled using CMFGEN models \citep{hillier98a}, those classified as O and B stars are handled using WM-Basic models \citep{pauldrach01a}, and the balance are treated using Kurucz atmospheres as catalogued by \citet{lejeune97a} \red{(referred to as the BaSeL libraries)}. Different combinations of the \red{BaSeL}, \citet{hillier98a}, and \citet{pauldrach01a} atmospheres are also possible.

\section{\red{Modeling Nebular Emission and Extinction}}
\label{app:nebuladust}

\red{Here we describe \slug's model for computing nebular emission and dust extinction.}

\subsection{\red{Nebular Continuum and Hydrogen Lines}}

\red{As described in the main text, the nebular emission rate can be written as $L_{\lambda,\mathrm{neb}} = \gamma_{\mathrm{neb}} \phi Q(\mathrm{H}^0) /\alpha^{\mathrm{(B)}}(T)$.} \Slug\ computes $\alpha^{\mathrm{(B)}}(T)$ using the analytic fit provided by \citet[his equation 14.6]{draine11a}, and adopts a fiducial value of $\phi$ following \citet{mckee97a}. However, the user is free to alter $\phi$. \red{Similarly, the temperature $T$ can either be user-specified as a fixed value, or can be set automatically from the tabulated \texttt{cloudy} data (see Appendix \ref{appsub:metlines}).}

\Slug\ computes the nebular emission coefficient as
\begin{eqnarray}
\gamma_{\lambda,\mathrm{neb}} & = &
\gamma_{\mathrm{ff}}^{(\mathrm{H})} + \gamma_{\mathrm{bf}}^{(\mathrm{H})} + \gamma_{\mathrm{2p}}^{(\mathrm{H})} 
+ \sum_{n<n'} \alpha_{nn'}^{\mathrm{eff,(B),(H)}} E_{nn'}^{(\mathrm{H})}
 \nonumber \\
& &
+  x_{\mathrm{He}} \gamma_{\mathrm{ff}}^{(\mathrm{He})} 
+ x_{\mathrm{He}} \gamma_{\mathrm{bf}}^{(\mathrm{He})}
+ \red{\sum_i \gamma_{i,\mathrm{line}}^{(\mathrm{M})}}.
\end{eqnarray}
The terms appearing in this equation are the helium abundance relative to hydrogen $x_{\mathrm{He}}$, the H$^+$ and He$^+$ free-free emission coefficients $\gamma_{\mathrm{ff}}^{(\mathrm{H})}$ and $\gamma_{\mathrm{ff}}^{(\mathrm{He})}$, the H and He bound-free emission coefficients $\gamma_{\mathrm{bf}}^{(\mathrm{H})}$ and $\gamma_{\mathrm{bf}}^{(\mathrm{He})}$, the H two-photon emission coefficient $\gamma_{\mathrm{2p}}^{(\mathrm{H})}$, the effective emission rates for H recombination lines $\alpha_{nn'}^{\mathrm{eff,(B),(H)}}$ corresponding to transitions between principal quantum numbers $n$ and $n'$, the energy differences $E_{nn'}$ between these two states, and \red{the emission coefficient for various metal lines (including He lines) $\gamma_{i,\mathrm{line}}^{(\mathrm{M})}$.}

\red{We compute each of these quantities as follows. For the H and He free-free emission coefficients,} $\gamma_{\mathrm{ff}}^{(\mathrm{H})}$ and $\gamma_{\mathrm{ff}}^{(\mathrm{He})}$, \red{we use} the analytic approximation to the free-free Gaunt factor given by \citet[his equation 10.8]{draine11a}. \red{We obtain the corresponding bound-free emission coefficients,} $\gamma_{\mathrm{bf}}^{(\mathrm{H})}$ and $\gamma_{\mathrm{bf}}^{(\mathrm{He})}$, by interpolating on the tables provided by \citet{ercolano06a}. \red{We obtain the effective case B recombination rate coefficients,} $\alpha_{nn'}^{\mathrm{eff,(B),(H)}}$, by interpolating on the tables provided by \citet{storey95a}. In practice, the sum includes all transitions for which the upper principal quantum number $n' \leq 25$.
We compute hydrogen two-photon emission via
\begin{equation}
\gamma_{\mathrm{2p}}^{(\mathrm{H})} = \frac{hc}{\lambda^3} I(\mathrm{H}^0) \alpha_{2s}^{\mathrm{eff,(B),(H)}} \frac{1}{1 + n_{\rm H}/n_{2s, \rm crit}} P_\nu,
\end{equation}
where $\alpha_{2s}^{\mathrm{eff,(B),(H)}}$ is the effective recombination rate to the $2s$ state of hydrogen in case B, interpolated from the tables of \citet{storey95a}, $n_{2s,\mathrm{crit}}$ is the critical density for the $2s-1s$ transition, and $P_\nu$ is the hydrogen two-photon frequency distribution, computed using the analytic approximation of \citet{nussbaumer84a}. The critical density in turn is computed as
\begin{equation}
n_{2s, \rm crit} = \frac{A_{2s-1s}}{q_{2s-2p,p} + (1+x_{\mathrm{He}}) q_{2s-2p,e}},
\end{equation}
where $A_{2s-1s} = 8.23$ s$^{-1}$ is the effective Einstein coefficient $A$ for the $2s-1s$ transition \citep[section 14.2.4]{draine11a}, and $q_{2s-2p,p}$ and $q_{2s-2p,e}$ are the rate coefficients for hydrogen $2s-2p$ transitions induced by collisions with free protons and free electrons, respectively. We take these values from \citet[table 4.10]{osterbrock89a}.

\subsection{\red{Non-Hydrogen Lines}}
\label{appsub:metlines}

\red{
Metal line emission is somewhat trickier to include. The emission coefficients for metal lines can vary over multiple orders of magnitude depending on H~\textsc{ii} region properties such as the metallicity, density, and ionization parameter, as well as the shape of the ionizing spectrum. Fully capturing this variation in a tabulated form suitable for fast computation is not feasible, so we make a number of assumptions to reduce the dimensionality of the problem. We consider only uniform-density H~\textsc{ii} regions with an inner wind bubble of negligible size -- in the terminology of \citet{yeh12a}, these are classical Str\"{o}mgren spheres as opposed to wind- or radiation-confined shells. To limit the number of possible spectra we consider, we also consider only spectral shapes corresponding to populations that fully sample the IMF. Thus while our fast estimate still captures changes in the strength of line emission induced by stochastic fluctuations in the overall ionizing luminosity, it does not capture the additional fluctuations that should result from the shape of the ionizing spectrum. A paper studying the importance of these secondary fluctuations is in preparation.
}

\red{
With these choices, the properties of the H~\textsc{ii} region are to good approximation fully characterized by three parameters: stellar population age, metallicity, and ionization parameter. To sample this parameter space, we use \texttt{cloudy} \citep{ferland13a} to compute the properties of H~\textsc{ii} regions on a grid specified as follows:
\begin{itemize}
\item We consider input spectra produced by using \slug~to calculate the spectrum of a $10^3$ $M_\odot$ mono-age stellar population with a \citet{chabrier05a} IMF and each of our available sets of tracks (see Appendix \ref{app:tracks}), at ages from $0-10$ Myr at intervals of $0.2$ Myr. We also consider the spectrum produced by continuous star formation at $\dot{M}_* = 10^{-3}$ $M_\odot$ yr$^{-1}$ over a 10 Myr interval. Note that the mass and star formation rate affect only the normalization of the spectrum, not its shape.
\item For each set of tracks, we set the H~\textsc{ii} region metallicity relative to Solar equal to that of the tracks. Formally, we adopt \texttt{cloudy}'s ``H~\textsc{ii} region" abundances case to set the abundance pattern, and then scale the abundances of all gas and grain components by the metallicity of the tracks relative to Solar.
\item The ionization parameter, which gives the photon to baryon ratio in the H~\textsc{ii} region, implicitly specifies the density. To complete this specification, we must choose where to measure the ionization parameter, since it will be high near the stars, and will fall off toward the H~\textsc{ii} region outer edge. To this end, we define our ionization parameter to be the volume-averaged value, which we compute in the approximation whereby the ionizing photon luminosity passing through the shell at radius $r$ is \citep[section 15.3]{draine11a}
\begin{equation}
Q(\mathrm{H}^{0}, r) = Q(\mathrm{H}^{0}) \left[1-\left(\frac{r}{r_s}\right)^3\right].
\end{equation}
Here $Q(\mathrm{H}^{0})$ is the ionizing luminosity of stars at $r=0$ and
\begin{equation}
r_s = \left( \frac{3Q(\mathrm{H}^{0})} {4\pi \alpha^{\mathrm{(B)}} n_{\mathrm{H}}^2}\right)^{1/3}
\end{equation}
is the classical Str\"omgren radius, which we compute using $\alpha^{\mathrm{(B)}}$ evaluated at a temperature of $10^4$ K. With these approximations, the volume-averaged ionization parameter is
\begin{eqnarray}
\langle\mathcal{U}\rangle 
& = & \frac{3}{4\pi r_s^3}\int_0^{r_s} 4\pi r^2 
\left(\frac{Q(\mathrm{H}^{0})}{4\pi r^2 c n_{\mathrm{H}}}\right)
\left[1-\left(\frac{r}{r_s}\right)^3\right]\, dr 
\nonumber \\
& = & \left[\frac{81 \left(\alpha^{\mathrm{(B)}}\right)^2 n_{\mathrm{H}} Q(\mathrm{H}^{0})}{288\pi c^3}\right]^{1/3}.
\label{eq:ionparam}
\end{eqnarray}
Thus a choice of $\langle\mathcal{U}\rangle$, together with the total ionizing luminosity determined from the \slug~calculation, implicitly sets the density $n_{\mathrm{H}}$ that we use in the \texttt{cloudy} calculation. Note that, as long as $n_{\mathrm{H}}$ is much smaller than the critical density of any of the important collisionally-excited lines, the exact value of $Q(\mathrm{H}^{0})$ and thus $n_{\mathrm{H}}$ changes only the absolute luminosities of the lines. The ratios of line luminosity to $Q(\mathrm{H}^{0})$, and thus the emission coefficients $\gamma^{(\mathrm{M})}$, depend only $\langle\,\mathcal{U}\rangle$. Our grid of \texttt{cloudy} calculations uses $\log\,\langle\mathcal{U}\rangle = -3$, $-2.5$, and $-2$, which spans the plausible observed range.
\end{itemize}
}

\red{After using \texttt{cloudy} to compute the properties of H~\textsc{ii} regions for each track set, age, and ionization parameter in our grid, we record the emission-weighted mean temperature $\langle T \rangle = \int n_{\mathrm{H}}^2 T \, dV / \int n_{\mathrm{H}}^2\, dV$, and the ratio $L_{\mathrm{line}} / Q(\mathrm{H}^0)$ for all non-hydrogen lines for which the ratio exceeds $10^{-20}$ erg photon$^{-1}$ at any age; for comparison, this ratio is $\approx 10^{-12}$ erg photon$^{-1}$ for bright lines such as H$\alpha$. This cut typically leaves $\sim 80$ significant lines. These data are stored in a tabular form. When a \slug~simulation is run, the user specifies an ionization parameter $\langle\mathcal{U}\rangle$. To compute nebular line emission, \slug~loads the tabulated data for the specified evolutionary tracks and $\langle\mathcal{U}\rangle$, and for each cluster or field star of known age interpolates on the grid of ages to determine $L_{\mathrm{line}} / Q(\mathrm{H}^0)$ and thus $\gamma^{(\mathrm{M})}$; field stars that are not being treated stochastically are handled using the values of $L_{\mathrm{line}} / Q(\mathrm{H}^0)$ tabulated for continuous star formation. At the user's option, this procedure can also be used to obtain a temperature $\langle T \rangle$, which in turn can be used in all the calculations of H and He continuum emission, and H recombination line emission.
}

\red{There is a final technical subtlety in \slug's treatment of nebular emission. The wavelength grid found in most broadband stellar atmosphere libraries, including those used by \slug, is too coarse to represent a nebular emission line. This leads to potential problems with the representation of such lines, and the calculation of photometry in narrowband filters targeting them (e.g., H$\alpha$ filters). To handle this issue, \slug~computes nebular spectra on a non-uniform grid in which extra wavelength resolution is added around the center of each line. The extra grid points make it possible to resolve the shape of the line (which we compute by adopting a Gaussian line shape with a fiducial width of 20 km s$^{-1}$), at least marginally. The grid resolution is high enough so that it is possible to compute numerical integrals on the spectrum and recover the correct bolometric luminosity of each line to high accuracy, so that photometric outputs including the nebular contribution can be computed correctly.}

\subsection{\red{Dust Extinction}}

\red{\Slug~parametrizes dust extinction via the $V$-band extinction, $A_V$. As noted in the main text, $A_V$ can either be a constant value, or can be specified as a PDF as described in Appendix \ref{app:pdfs}. In the latter case, every star cluster has its own extinction, so a range of extinctions are present. Once a value of $A_V$ is specified,} \slug\ computes the wavelength-dependent extinction from a user-specified extinction law. The extinction curves that ship with the current version of the code are as follows:
\begin{itemize}
\item a Milky Way extinction curve, consisting of optical and UV extinctions taken from \citet{fitzpatrick99a}, and IR extinctions taken from \citet{landini84a}, with the two parts combined by D.~Calzetti (priv.~comm., 2014).
\item a Large Magellanic Cloud extinction curve, taken from the same source as the Milky Way curve
\item a Small Magellanic Cloud extinction curve, taken from \citet{bouchet85a}
\item a ``starburst" extinction curve, taken from \citet{calzetti00a}.
\end{itemize}

\section{Software Notes}
\label{app:software}

Full details regarding the code implementation are included in the \slug\ documentation, but we include in this \red{Appendix some details that are of general interest}. First, \red{\slug, \cloudyslug, and various related software packages are fully parallelized for multi-core environments.} Second, the \slug\ package includes a python helper library, \slugpy, that is capable of reading and manipulating \slug\ outputs\red{, and which integrates with \cloudyslug, \sfrslug, and \cs}. In addition to more mundane data processing tasks, the library supports ancillary computations such as convolving spectra with additional filters and converting data between photometric systems. The \slugpy\ library is also fully integrated with all the tools described below. FITS file handling capabilities in \slugpy\ are provided through \texttt{astropy} \citep{astropy-collaboration13a}. Third, the code is highly modular so that it is easy to add additional options. Extinction curves and photometric filters are specified using extremely simple text file formats, so adding additional filters or extinction curves is simply a matter of placing additional text files in the relevant directories. Similarly, probability distributions describing IMFs, CMFs, SFHs, etc., are also specified via simple text files, so that additional choices can be added without needing to touch the source code in any way.

\red{The numerical implementation used in \bp~requires particular attention, since having a fast implementation is critical for the code's utility.} Since we have written the joint and marginal posterior probability distributions of the physical variables in terms of kernel density estimates (equations \ref{eq:pdf} and \ref{eq:mpdf}), we can perform numerical evaluation using standard fast methods. In \bp, numerical evaluation proceeds in a number of steps. After reading in the library of simulations, we first compute the weights from the user-specified prior probability distributions and sampling densities (\autoref{eq:weight}). We then store the sample points and weights in a $k-$dimensional (KD) tree structure. The bandwidth we choose for the kernel density estimation must be chosen appropriately for the input library of models, and for the underlying distribution they are modeling. There is no completely general procedure for making a ``good" choice for the bandwidth, so bandwidth selection is generally best done by hand.

Once the bandwidth has been chosen, we can evaluate the joint and marginal posterior PDFs to any desired level of accuracy by using the KD tree structure to avoid examining parts of the simulation library are not relevant for any particular set of observations. As a result, once the tree has been constructed, the formal order of the algorithm for evaluating either the joint or marginal PDF using a library of $N_{\rm lib}$ simulations is only $\log N_{\rm lib}$, and in practice evaluations of the marginal PDF over relatively fine grids of points can be accomplished in well under a second on a single CPU, even for 5-band photometry and libraries of many millions of simulations. In addition to evaluating the joint or marginal PDF directly on a grid of sample points, if we are interested in the joint PDF of a relatively large number of physical variables, it may be desirable to use a Markov Chain Monte Carlo (MCMC) method to \red{explore the shape of the posterior PDF}. \Bp\ includes an interface to the MCMC code \texttt{emcee} \citep{foreman-mackey13a}, allowing transparent use of an MCMC technique as well as direct evaluation on a grid. However, if we are interested in the marginal PDFs only of one or two variables at time (for example the marginal PDF of star cluster mass or star cluster age, or their joint distribution), it is almost always faster to use equation (\ref{eq:mpdf}) to evaluate this directly than to resort to MCMC. The ability to generate marginal posterior PDFs directly represents a significant advantage to our method, since this is often the quantity of greatest interest.

\end{appendix}

\label{lastpage}

\end{document}